\documentclass[twocolumn]{aastex62}

\graphicspath{{./}{figures/}}
\usepackage{gensymb}
\usepackage{amsmath}
\usepackage{appendix}
\usepackage{flafter}
\usepackage{xspace}

\newcommand{\sub}{MAXI~J1820+070\xspace}
\newcommand{\Ni}{\emph{NICER}\xspace}

\newcommand{\hx}{Insight-\emph{HXMT}\xspace}
\newcommand{\hbAppendixPrefix}{A}

\shorttitle{RMS-flux slope in MAXI~J1820+070}
\shortauthors{Wang et al.}

\begin{document}

\title{
\Large \textbf{RMS-flux slope in MAXI~J1820+070: a measure of the disk-corona coupling}}

\author{Yanan Wang} 
\affil{Key Laboratory of Optical Astronomy, National Astronomical Observatories, Chinese Academy of Sciences, Beijing 100101, China}
\altaffiliation{E-mail: wangyn@bao.ac.cn}
\author{Shuang-Nan Zhang} 
\affiliation{Key Laboratory for Particle Astrophysics, Institute of High Energy Physics, Chinese Academy of Sciences, 19B Yuquan Road, Beijing 100049, China}
\affiliation{University of Chinese Academy of Sciences, Chinese Academy of Sciences, Beijing 100049, China}

\date{Accepted ? December ?. Received ? December ?; in original form ? December ?}

\begin{abstract}
Linear RMS-flux relation has been well established in different spectral states of all accreting systems. In this work, we study the evolution of the frequency-dependent RMS-flux relation of MAXI~J1820+070 during the initial decaying phase of the 2018 outburst with {\hx} over a broad energy range 1--150\,keV. 
As the flux decreases, we first observe a linear RMS-flux relation at frequencies from 2\,mHz to 10\,Hz, while such a relation breaks at varying times for different energies, leading to a substantial reduction in the slope.
Moreover, we find that the low-frequency variability exhibits the highest sensitivity to the break, which occurs prior to the hard-to-hard state transition time determined through time-averaged spectroscopy, and the time deviation increases with energy. The overall evolution of the RMS-flux slope and intercept suggests the presence of a two-component Comptonization system. One component is radially extended, explaining the strong disk-corona coupling before the break, while the other component extends vertically, contributing to the reduction of the disk-corona coupling after the break. A further vertical expansion of the latter component is required to accommodate the dynamic evolution observed in the RMS-flux slope.
In conclusion, we suggest that the RMS-flux slope in 1--150\,keV band can be employed as an indicator of the disk-corona coupling and the hard-to-hard state transition in \sub could be partially driven by the changes in the corona geometry. 

\end{abstract}


\section{Introduction} \label{intro}

Based on the timing and spectral properties, two spectral states, soft and hard states, have been classified in black hole X-ray binaries (BHXRBs) to quantify the evolution of outbursts \citep{Tanaka1995,Klis1995,Mendez1997a,Remillard2006}. In the soft state, the spectrum is dominated by a multi-temperature blackbody component and the power spectrum is flat and featureless; while in the hard state, the spectrum is Comptonization dominated and the power spectrum is complex, described by multiple Lorentzian components with different centroid frequencies. Since the two states have different characteristics, the values of the spectral and timing parameters vary considerably from one state to another, and there are some phenomena, e.g. quasi-periodic oscillations (QPOs), jets and disk winds, that are exclusively present in certain states.
To distinguish between the two states, both spectral and timing properties should be considered. However, mechanisms driving state transitions in accreting BHs remain debated. 

One association that has been extensively studied in accretion systems is the linear root mean square (RMS)-flux relation (e.g. \citealt{Gleissner2004,Uttley2005,Heil2012}), which states that a greater mean source flux is associated with a proportionally larger absolute rms amplitude variability.
Such a relationship, together with non-linear variability and a lognormal flux distribution, suggests the variability is most likely the result of fluctuations generated at different radii propagating within an accretion disk \citep{Lyubarskii1997,Churazov2001,Gleissner2004,Uttley2005,Mushtukov2018}.
Additionally, the RMS-flux relation or a rms-intensity diagram (RID) has been tested as a good tracer of accretion states throughout outbursts \citep{Munoz2011}. 

The BH transient MAXI~J1820+070 was first discovered on 2018 March 11 in X-rays with the {\emph Monitor of All-sky X-ray Image} (\emph{MAXI}, \citealt{Matsuoka2009}) and has subsequently been observed from radio to hard X-rays by ground-based and space telescopes. It was first classified as a BH candidate based on its multi-wavelength properties, and then recognized as a BH transient after precise measurements of its BH mass and distance to the system ($M_{\rm BH}=9.5\pm1.4~M_{\rm \odot}$ and $D=3$\,kpc, \citealt{Atri2020}). 


The co-evolution of the disk and corona during the initial bright outburst of \sub has been extensively investigated through both timing and spectral analyses (e.g. \citealt{Kara2019,Wang2020,DeMarco2021,Zdziarski2021,Kawamura2023}). However, the geometry and location of these two emitting regions remain inconclusive and controversial.
For instance, in their study, \cite{Kara2019} propose the presence of a vertically extended corona that contracts as the outburst declines, with the requirement for a constant disk inner radius. Conversely, \cite{DeMarco2021}, analyzing the same data, report that the disk inner radius was initially truncated at the beginning of the outburst and subsequently moved inwards until it reached the innermost stable circular orbit.
Furthermore, both \cite{Zdziarski2021} and \cite{Kawamura2023} support a truncated disk scenario and put forth more complex coronal geometries. The former proposes two Comptonization regions: a soft one located above the disk and a hard one situated between the disk and the central object. The latter also suggests two Comptonization regions, both positioned between the disk and the central object, with the harder one being closer to the central object. 

Additionally, \cite{Wang2020} proposed that \sub experienced a failed outburst, marked by a hard-to-hard state transition around MJD 58257, during the initial outburst period between MJD~58190 and 58290. This finding is supported by the discernible patterns of evolution in hardness ratios, fractional rms, and time lags derived from timing analysis. Moreover, the evolution of the disk blackbody temperature and the spectral photon index, as deduced from spectral analysis, substantiate the finding. 
Nonetheless, in other BHXRBs undergoing failed or hard-only outbursts, the hardness ratio normally shows a monotonic increase as the flux decreases (e.g. \citealt{Capitanio2009,Bassi2019}). Consequently, the observed phenomena in \sub set this hard-to-hard transition apart from the canonical hard-to-soft transition, and the associated outburst also differs from other failed or hard-only outbursts. Similarly, \cite{Stiele2020} reach a comparable conclusion by examining the evolution of QPO frequencies using a distinct dataset.

In this work, we leverage the RMS-flux relation within Fourier frequencies ranging from 2\,mHz to 10\,Hz and across energies spanning 1 to 150\,keV. Our objective is to impose further constraints on the coronal geometry and investigate its correlation with the state transition of \sub during the initial decaying phase, spanning from MJD 58197 to 58288, utilizing \hx data.


\begin{figure*}
\centering
\mbox{\includegraphics[width=0.5\linewidth]{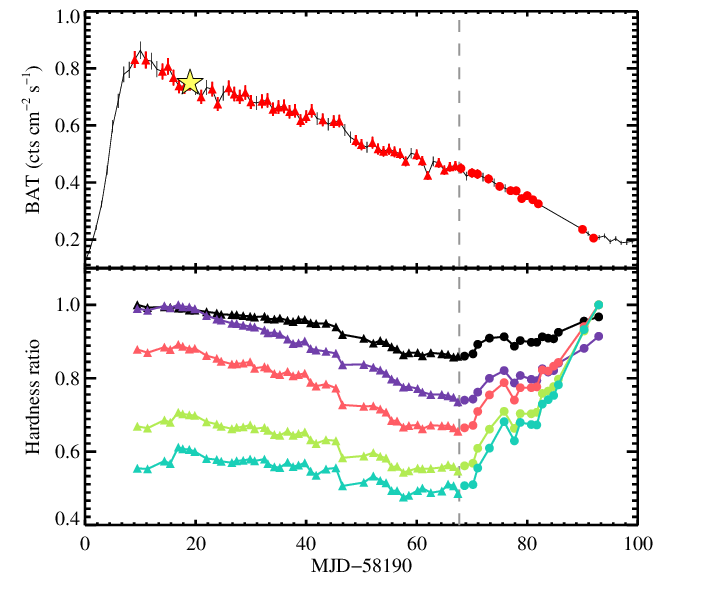}}
\hspace{-1cm}
\mbox{\includegraphics[width=0.5\linewidth]{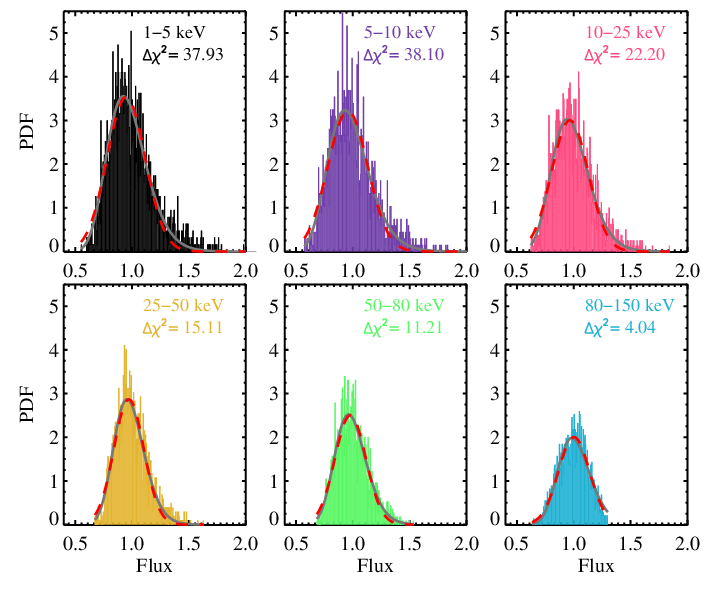}}
\vspace{-0.1cm}
\caption{\textbf{Left:} The long-term BAT (15--50\,keV) lightcurve and the hardness ratio of count rates in the energy bands 5--10\,keV, 10--25\,keV, 25--50\,keV, 50--80\,keV, 80--150\,keV, respectively, with respect to the 1--5\,keV band. \textbf{The red triangles and circles in the upper panel indicate the simultaneous \hx observations analyzed in this work.}
The hardness ratio decreases as the energies decreases. The yellow star marks the time of the observation, serving as an illustrative example for the PDF of flux in the right panels.
\textbf{Right:} Examples of PDF of flux (in units of counts per second) at different energies. The gray solid and red dashed lines indicate the lognormal and Gaussian fits, respectively. The legends $\Delta \chi^2$ represent the disparity in the $\chi^2$ values for the Gaussian and the lognormal fits, and a positive value indicates that the latter yields a smaller $\chi^2$.} 
\label{fig:lognormal}\label{fig:lc_PDF}
\end{figure*}

\section{Observations and data reduction}
The entire outburst of {\sub} has been observed with {\hx} between March and October 2018.
Owing to the broad energy range and high time resolution of {\hx}, we study the frequency dependent RMS-flux relation of {\sub}, up to 150~keV, of 63 observations from MJD~58197 to 58288. The data were reduced as in \cite{Wang2020}.

For the computation of the rms, we followed the steps in \cite{Heil2012} to calculate the rms in each continuous 500\,s segment with a time bin of 0.005\,s; the rms is determined by computing a periodogram (in fractional rms normalization) for each segment, and is then estimated by integrating the averaged power spectrum over the 2--10\,mHz, 0.01--1\,Hz and 1--10\,Hz frequency ranges. 
We used the Python-based package Stingray\footnote{https://github.com/StingraySoftware/stingray} to perform the above calculation, in which the Poisson noise level at 2 in Leahy normalization is subtracted from each periodogram.
The uncertainties on the rms estimates are also computed as described by \cite{Heil2012}. We computed the rms for the data in six energy bands, i.e. 1--5\,keV, 5--10\,keV, 10--25\,keV, 25--50\,keV, 50-80\,keV and 80-150\,keV, respectively, and multiplied the fractional rms with the mean count rate in each energy bands to obtain the absolute rms.


\begin{figure*}  
\centering
\mbox{\includegraphics[width=3.3cm]{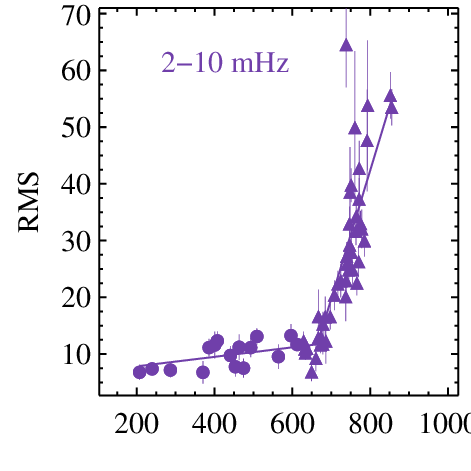}}
\hspace{-0.6cm}
\vspace{-0.2cm}
\mbox{\includegraphics[width=3.3cm]{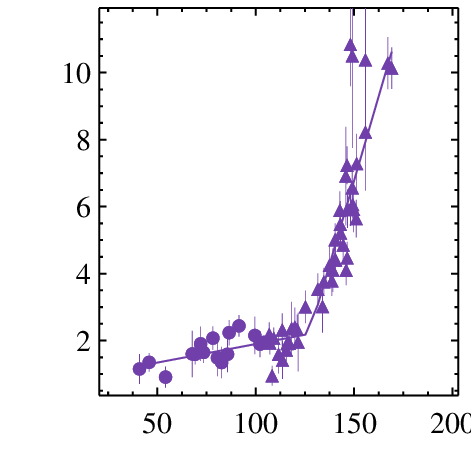}}
\hspace{-0.6cm}
\mbox{\includegraphics[width=3.3cm]{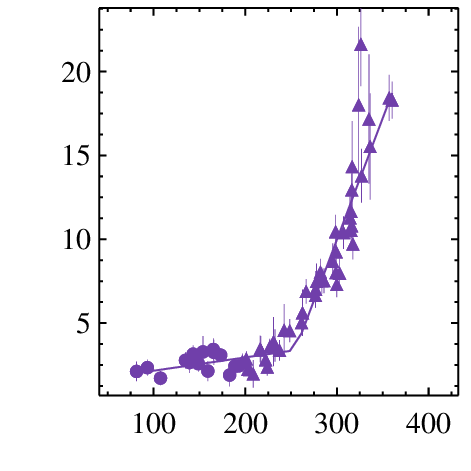}}
\hspace{-0.6cm}
\mbox{\includegraphics[width=3.3cm]{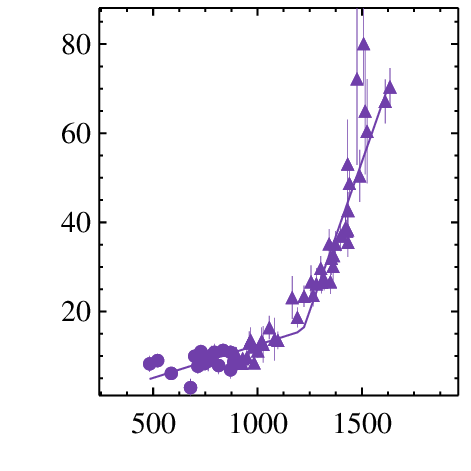}}
\hspace{-0.6cm}
\mbox{\includegraphics[width=3.3cm]{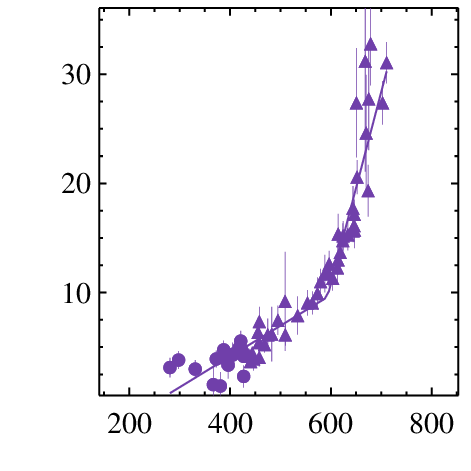}}
\hspace{-0.6cm}
\mbox{\includegraphics[width=3.3cm]{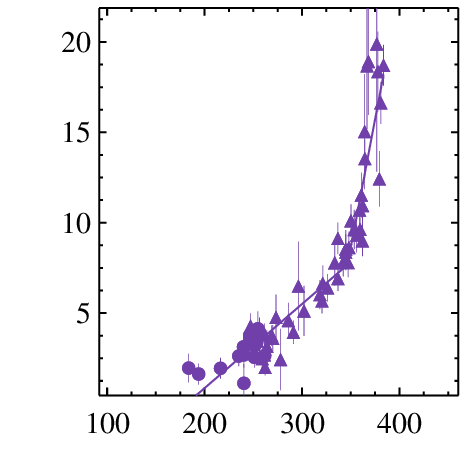}}
\vspace{-0.2cm}
\mbox{\includegraphics[width=3.3cm]{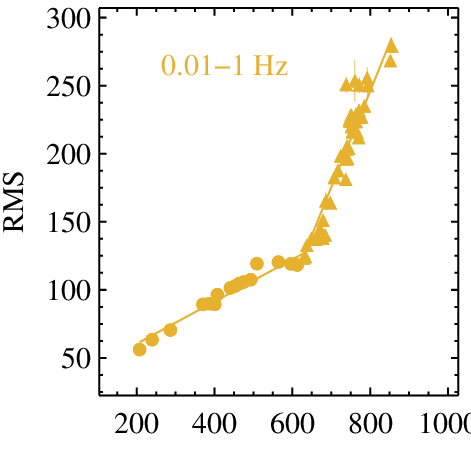}}
\hspace{-0.6cm}
\mbox{\includegraphics[width=3.3cm]{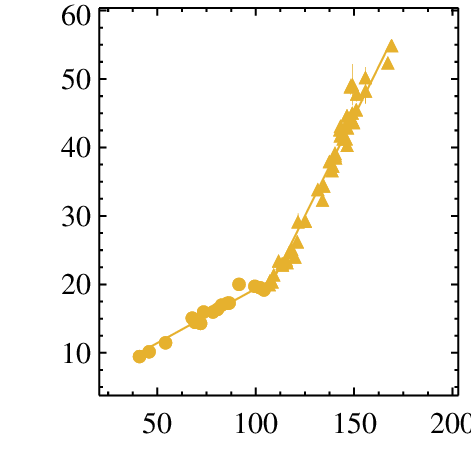}}
\hspace{-0.6cm}
\mbox{\includegraphics[width=3.3cm]{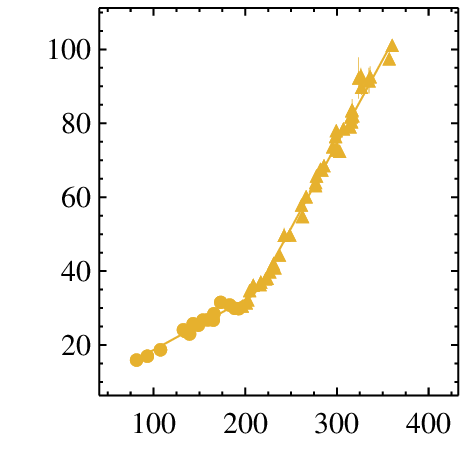}}
\hspace{-0.6cm}
\mbox{\includegraphics[width=3.3cm]{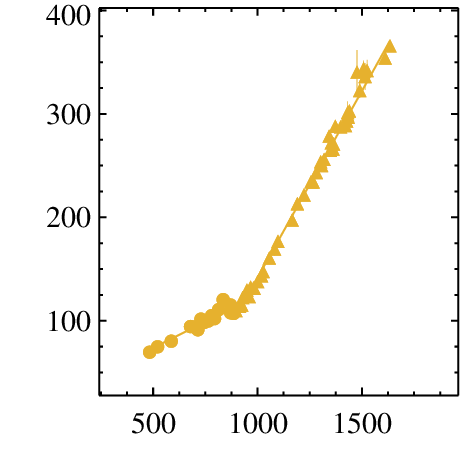}}
\hspace{-0.6cm}
\mbox{\includegraphics[width=3.3cm]{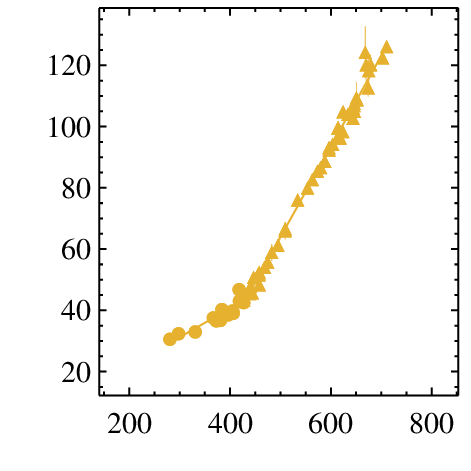}}
\hspace{-0.6cm}
\mbox{\includegraphics[width=3.3cm]{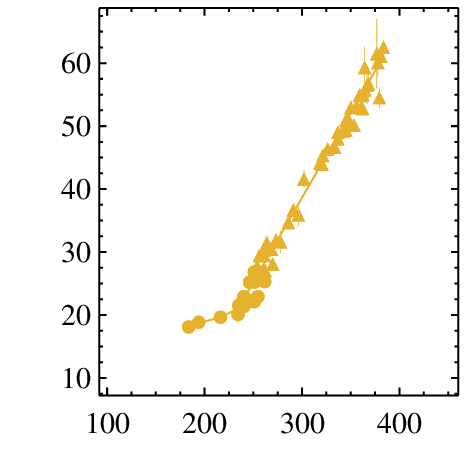}}
\mbox{\includegraphics[width=3.3cm]{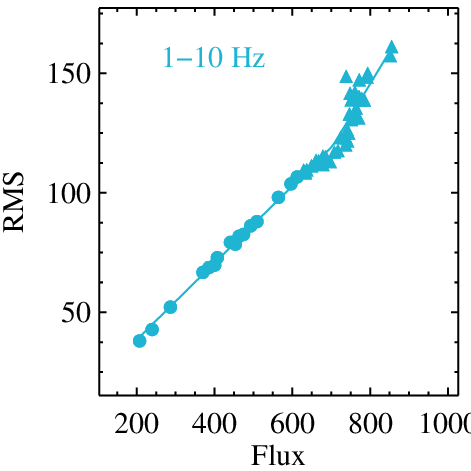}}
\hspace{-0.6cm}
\mbox{\includegraphics[width=3.3cm]{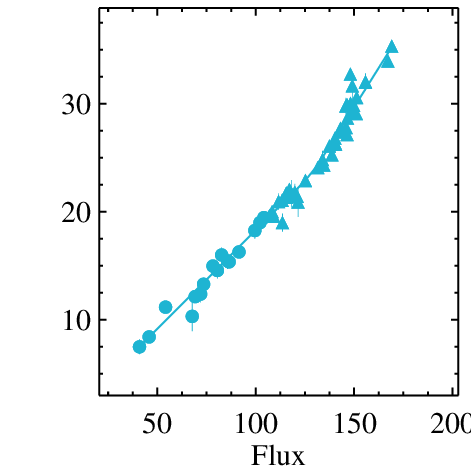}}
\hspace{-0.6cm}
\mbox{\includegraphics[width=3.3cm]{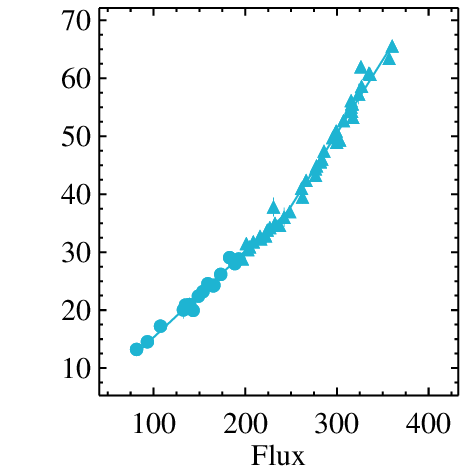}}
\hspace{-0.6cm}
\mbox{\includegraphics[width=3.3cm]{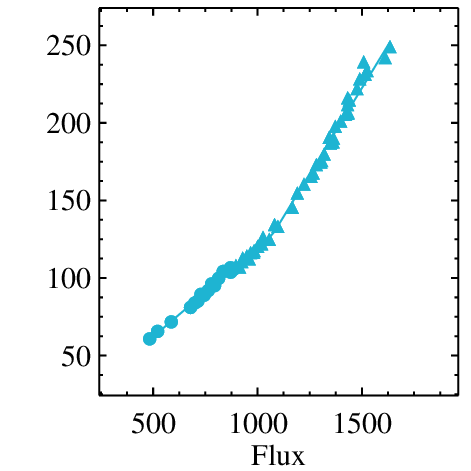}}
\hspace{-0.6cm}
\mbox{\includegraphics[width=3.3cm]{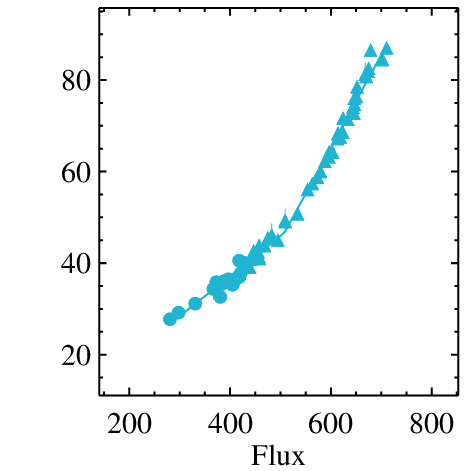}}
\hspace{-0.6cm}
\mbox{\includegraphics[width=3.3cm]{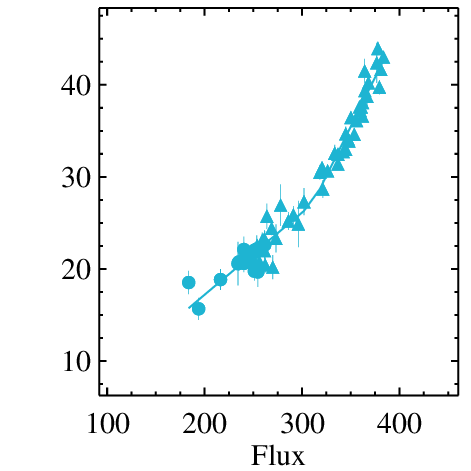}}
\vspace{-0.1cm}
\caption{Energy-dependent RMS-flux relation in {\sub} in three frequency ranges. The triangles and dots correspond to epochs~1 and 2, respectively, as defined in \protect\cite{Wang2020}. Each curve is fitted with a broken line.
}
\label{fig:rms_flux}
\end{figure*}

\begin{figure*}  
\centering
\mbox{\includegraphics[width=3.3cm]{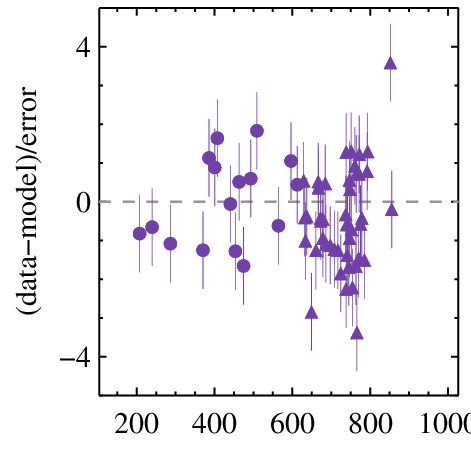}}
\hspace{-0.6cm}
\vspace{-0.2cm}
\mbox{\includegraphics[width=3.3cm]{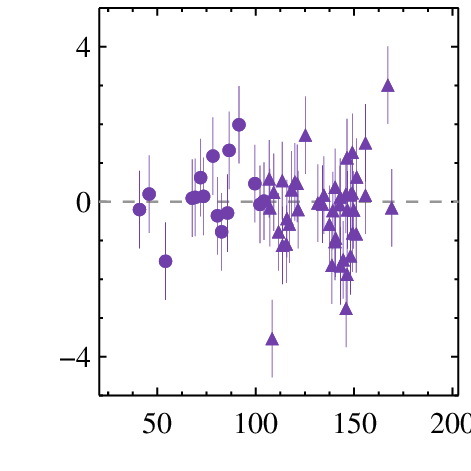}}
\hspace{-0.6cm}
\mbox{\includegraphics[width=3.3cm]{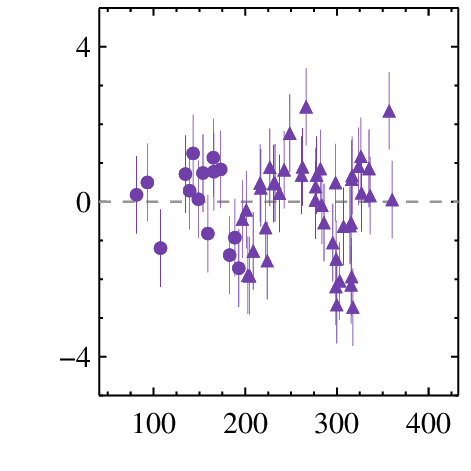}}
\hspace{-0.6cm}
\mbox{\includegraphics[width=3.3cm]{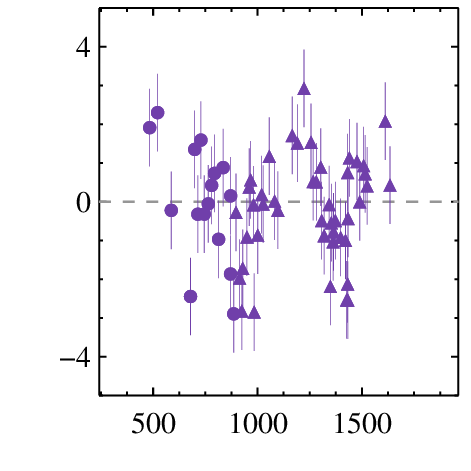}}
\hspace{-0.6cm}
\mbox{\includegraphics[width=3.3cm]{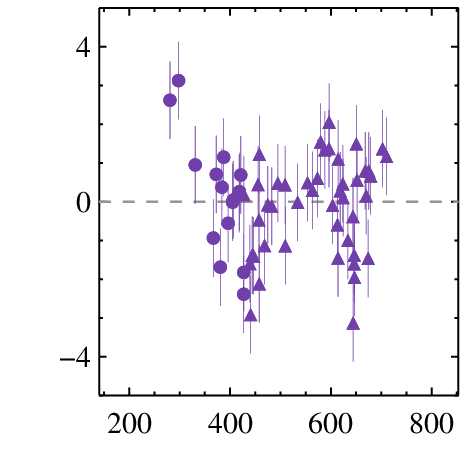}}
\hspace{-0.6cm}
\mbox{\includegraphics[width=3.3cm]{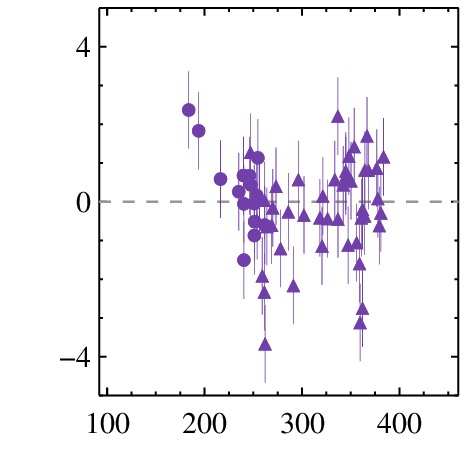}}
\vspace{-0.2cm}
\mbox{\includegraphics[width=3.3cm]{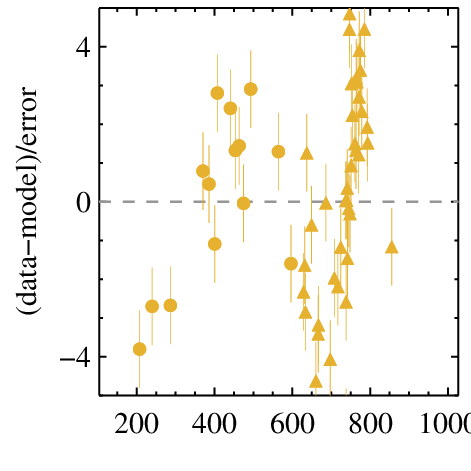}}
\hspace{-0.6cm}
\mbox{\includegraphics[width=3.3cm]{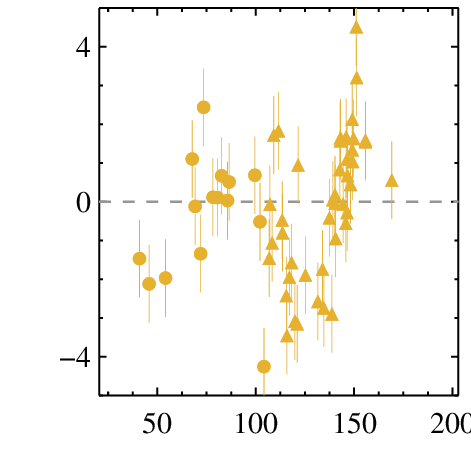}}
\hspace{-0.6cm}
\mbox{\includegraphics[width=3.3cm]{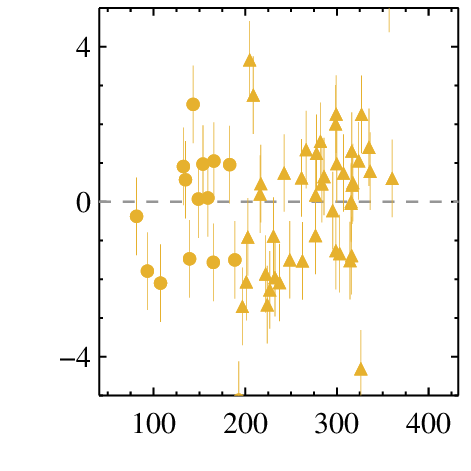}}
\hspace{-0.6cm}
\mbox{\includegraphics[width=3.3cm]{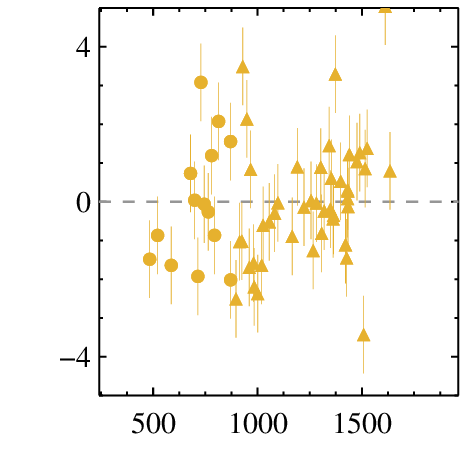}}
\hspace{-0.6cm}
\mbox{\includegraphics[width=3.3cm]{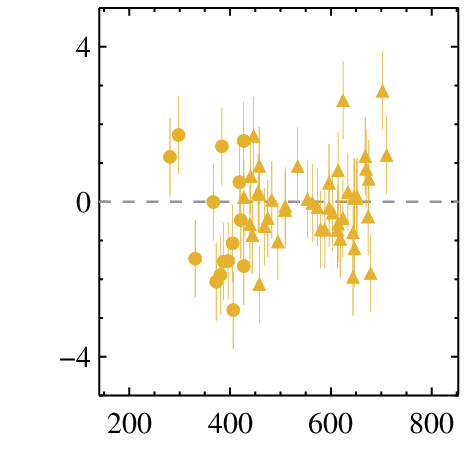}}
\hspace{-0.6cm}
\mbox{\includegraphics[width=3.3cm]{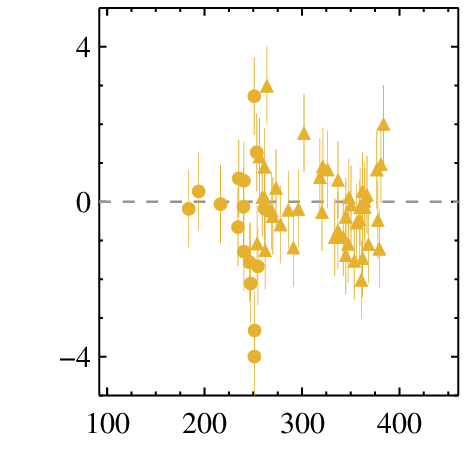}}
\mbox{\includegraphics[width=3.3cm]{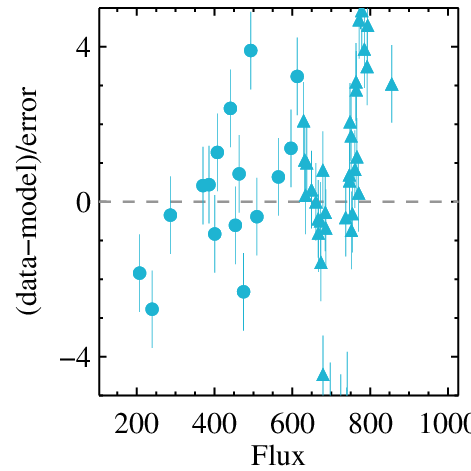}}
\hspace{-0.6cm}
\mbox{\includegraphics[width=3.3cm]{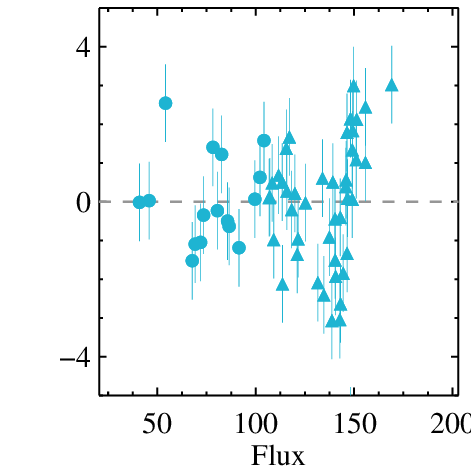}}
\hspace{-0.6cm}
\mbox{\includegraphics[width=3.3cm]{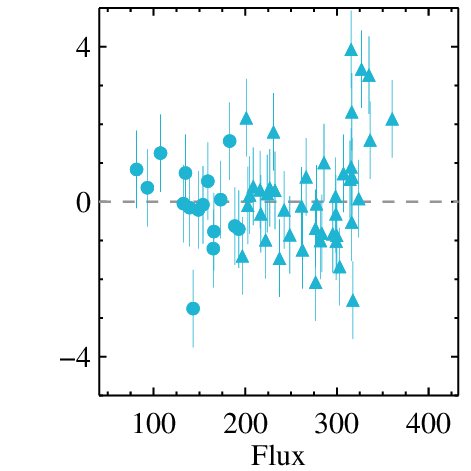}}
\hspace{-0.6cm}
\mbox{\includegraphics[width=3.3cm]{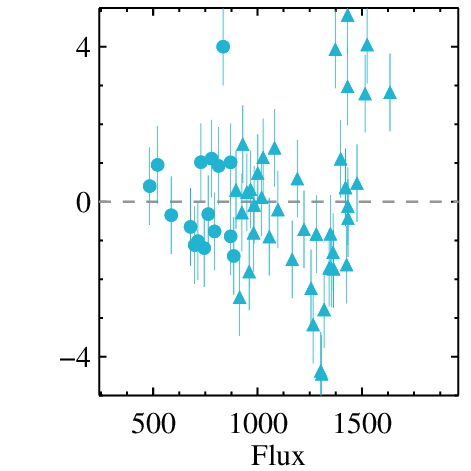}}
\hspace{-0.6cm}
\mbox{\includegraphics[width=3.3cm]{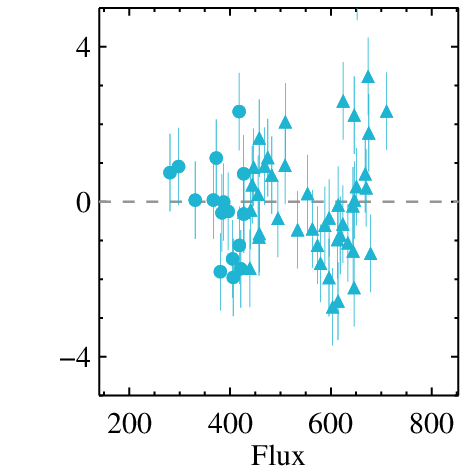}}
\hspace{-0.6cm}
\mbox{\includegraphics[width=3.3cm]{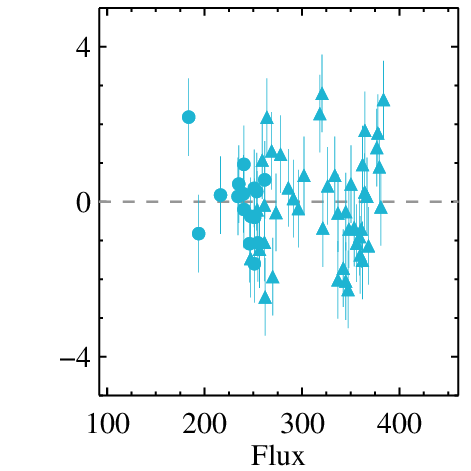}}
\vspace{-0.1cm}
\caption{Residuals of the fit to the RMS-flux relation. 
}
\label{fig:rms_flux_resi}
\end{figure*}

\section{Results}\label{sec:result_rms_flux}
We show the long-term BAT lightcurve and the hardness ratio of count rates between different energy bands of \hx in the left panels of Fig.~\ref{fig:lc_PDF}. The studied period of this work is denoted by triangles and dots. Where the two symbols connect signify the hard-to-hard transition, as proposed by \cite{Wang2020}.
Examples of the lightcurve and the corresponding periodograms of the studied data have been presented in Fig.~1 of \cite{Wang2020}. 
We plot the probability density distribution (PDF) of the flux of the same lightcurve in different energy bands in the right panels of Fig.~\ref{fig:lognormal}.
We performed a fitting of the PDF using either a lognormal or a Gaussian function. For equivalent degrees of freedom (307, 287, 242, 168, 168, and 131) in each energy band, the lognormal fit consistently provides a better fit, as indicated by smaller $\chi^2$ values (specifically, $\chi^2$ = 21.8, 28.0, 18.7, 9.6, 9.3, and 5.4) in comparison to the Gaussian fit ($\chi^2$ = 59.7, 66.1, 40.9, 24.8, 20.5, and 9.43).

\begin{figure}
\centering
\mbox{\includegraphics[width=1\linewidth]{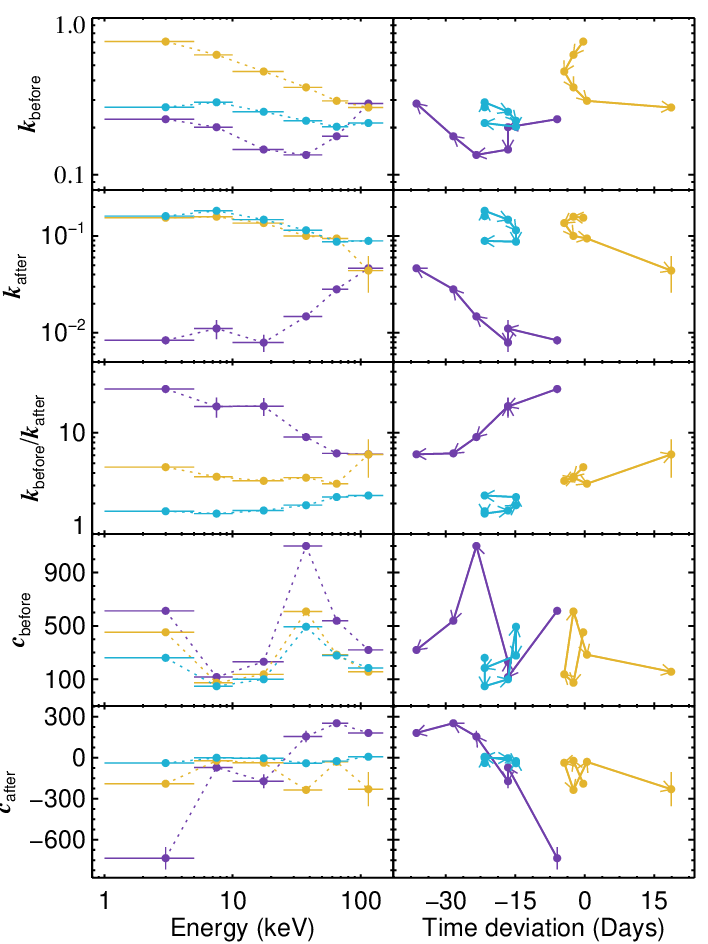}}
\caption{Rms-flux slopes/intercepts on the flux axis and their ratio against energy and time deviation for different frequencies. } 
\label{fig:slope_ene}
\end{figure}

We further plot the RMS-flux relation in Fig.~\ref{fig:rms_flux}, where the triangles and dots represent the data before and after MJD~58257 when the hard-to-hard state transition occurred as reported in several works (e.g. \citealt{Wang2020,Ma2021,You2021}).
Fig.~\ref{fig:rms_flux} shows that as the flux decreases, the RMS-flux curve deviates from a linear relation for all the selected frequencies, and the deviation is larger for lower frequencies. 
We suspect that such a deviation is related to the state transition. In fact, each complete RMS-flux curve can be characterized by a discontinuous line, but the location of this discontinuity appears to differ from where the triangles and dots intersect. 
To quantitatively describe the discontinuity, we fitted each curve with a broken line, which is expressed as:
\begin{equation}
\small
\sigma(t)=\left\{
\begin{array}{ll}
a+k_{\rm after}*b+k_{\rm before}*[F(t)-b], {t \leq t_{\rm break}}\\
a + k_{\rm after}*F(t), {t > t_{\rm break}},\\
\end{array}
\right.
\label{equ:flux_rms}
\end{equation}   
where $\sigma(t)$ and $F(t)$ is the RMS and the flux at time $t$, $a$ and $b$ are two constants, $k_{\rm before}$ and $k_{\rm after}$ are the slopes of the RMS-flux relation before and after the break, respectively. Hence the intercept of the RMS-flux relation on the flux axis for the two segments is $c_{\rm before}=b-(a+k_{\rm after}*b)/k_{\rm before}$ and $c_{\rm after}=-a/k_{\rm after}$. The best-fitting broken lines are added to Figs.~\ref{fig:rms_flux}. 
Furthermore, we show the slopes and intercepts against energy for different frequencies in the left panels of Figs.~\ref{fig:slope_ene}. 

The RMS-flux slope equals to the fractional rms of the Fourier power in the selected frequency range. As illustrated in Fig.~\ref{fig:slope_ene}, the values of $k_{\rm before}$ are higher than those of $k_{\rm after}$, and the 2--10\,mHz slope reduces most significantly for over one magnitude.
Moreover, regardless before or after the break, the 0.01--1\,Hz and 1--10\,Hz slopes decrease with energy; at the same time, the 2--10\,mHz slope first either decreases or remains constant with energy and then increases with energy above $\sim$25\,keV.

It is evident that the line break is not aligned with the connection between the triangles and the dots, suggesting that the state transition observed in the Fourier domain diverges from that inferred from spectroscopy.
To show the degree of deviation of the break time from the spectral transition time, we plot the slopes as a function of time in the right panels of Fig.~\ref{fig:slope_ene} in which the arrows are applied to point from low to high energy.  
As illustrated in the figure, for 2--10\,mHz the time deviation increases with energy, ranging between 6.0 and 36.3\,days, meaning that the break occurs the earliest at the highest energy; for 1--10\,Hz the time deviation exhibits a more focused pattern, occurring roughly 14.9--21.6\,days before the transition time; for 0.01--1\,Hz the time deviation varies from -4.4 to 18.6\,days, with the positive value distinguishing it from the patterns seen at other frequencies.

\begin{figure}
\centering
\mbox{\includegraphics[width=0.9\linewidth]{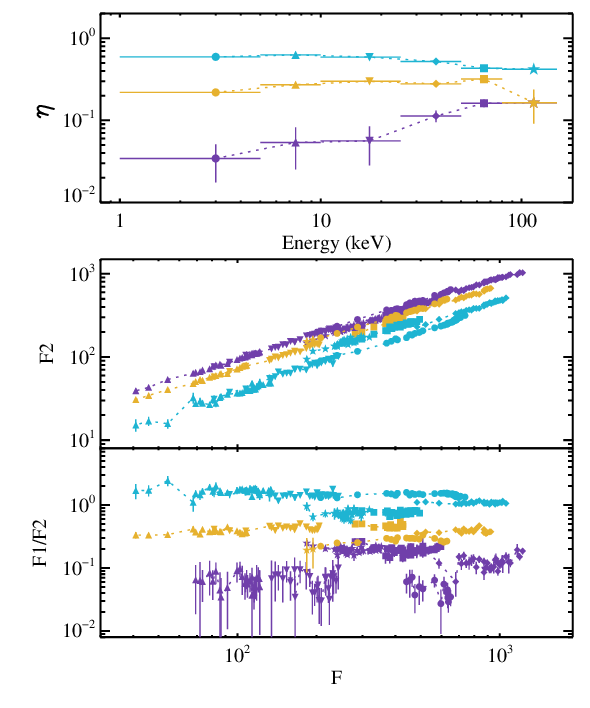}}
\vspace{-0.2cm}
\caption{\textbf{Upper:} $\eta$ against energy. \textbf{Lower:} $F_2$ and $F_1$/$F_2$ against $F$, respectively. Each symbol corresponds to a specific energy band, as illustrated in the upper panel.} 
\label{fig:alpha_f2}
\end{figure}

In terms of the intercept on the flux axis, prior to the break, the intercept-energy curve exhibits a similar trend across frequencies. The intercept represents the flux at which RMS equals zero and thus a positive value indicates the presence of a constant component within the system, devoid of variability. 
However, after the break, the intercept is negative in most cases. This result may initially appear non-physical as a flux would never be negative. To resolve this issue, we introduce an additional contributor, $F_{2}$, with a constant RMS to the emission. In this scenario, the total emission after the break consists of two components, $F_{1}$ donating the same linear RMS-flux relation as before the break and $F_{2}$ contributing a constant RMS. Hence, $F_{1}$ and $F_{2}$ are expressed as follows:
\begin{align}
F(t)&=F_{1}(t)+F_{2}(t) \label{equ:2},\\
F_{1}(t)&=F(t)*\eta(t) \label{equ:3},\\
\sigma(t)&=k_{\rm before}*F_{1}(t)+a \label{equ:4},
\end{align}
where $F(t)$ and $\sigma(t)$ represent the observed flux and RMS when $t>t_{\rm break}$, $k_{\rm before}$ and $a$ are adopted from Equation~\ref{equ:flux_rms}, and $\eta(t)$ denotes the ratio of $F_{1}(t)$ to $F(t)$. Following our calculations, we determine the value of $\eta$ and present the RMS as a function of $\eta$ in Fig.~\ref{fig:rms_eta}, which shows that $\eta$ rises with increasing frequencies. However, at higher frequencies, it remains independent of RMS, whereas at lower frequencies (i.e. 2-10\,mHz) and lower energies (i.e. below 50\,keV), it exhibits a proportional relationship with RMS. We further plot the average value of $\eta$ vs energy at each energy band, and $F_2$ and $F_1$/$F_2$ vs $F$ in Fig.~\ref{fig:alpha_f2}. Similar to the $k_{\rm after}$-energy curve, $\eta$ increases with energy at lower frequencies and slightly decreases with energy at higher frequencies. Furthermore, $F_1$ and $F_2$ both increase as $F$ increases, maintaining a constant energy-dependent ratio.


\section{Discussion}
We studied the evolution of the RMS-flux relation at different frequencies/energies of the black hole transient {\sub} in the decaying phase of the 2018 outburst while a hard-to-hard transition occurs. 
Four main results are obtained: i) the absolute RMS is linearly correlated with the flux with a break occurring at different times for different energies; ii) the RMS-flux slope significantly decreases after the break, and the low-frequency one shows the most dramatic reduction; iii) the time of break onset is energy-dependent, with the low-frequency break occurring earliest at the highest energies; 
iv) two components are required to account for the RMS-flux relation: before the break, one contributes to a steep linear RMS-flux relation, while the other maintains a constant flux without any fluctuations; after the break, one component still exhibits a linear RMS-flux relation with the same slope as before the break, whereas the other component remains constant in its RMS.
We discuss the implication of these results and interpret them within the framework of the evolution of the mass accretion rate and the geometry of the Comptonization region.




\subsection{Previous studies on RMS-flux relation}
Regardless of time-scales, accretion-induced variability has exhibited two common properties, one is a lognormal distribution of fluxes and another one is a linear RMS-flux relation. Such properties suggest this variability process to be multiplicative and thus it could be driven by accretion rate variations produced at different radii and/or result from variations in multiple dependent emitting regions \citep{Lyubarskii1997,Churazov2001,Uttley2005,Mushtukov2018}.

\cite{Gleissner2004} extended the study of the RMS-flux relation in the frequency range of 1--32\,Hz to different spectral states of Cyg~X--1. They found that the relation remains to be linear in all the spectral states but the RMS-flux slope in the soft and intermediate states appears to be shallower than in the hard state, with a difference of less than a factor of 2 between the soft and hard states. 
Unlike most BHXRBs, \sub did not experience a canonical state transition but presented a failed state transition in the studied period. If we assume the period associated with the rise/fall of the photon index as the period approaching to/departing from a soft state, the evolution of the 1--10\,Hz slope is consistent with their result. However, the evolution of the RMS-flux slope at lower frequencies has yet to be explored.

\subsection{Characteristics of the studied data}
\begin{figure*}
\centering
\mbox{\includegraphics[width=0.3\linewidth]{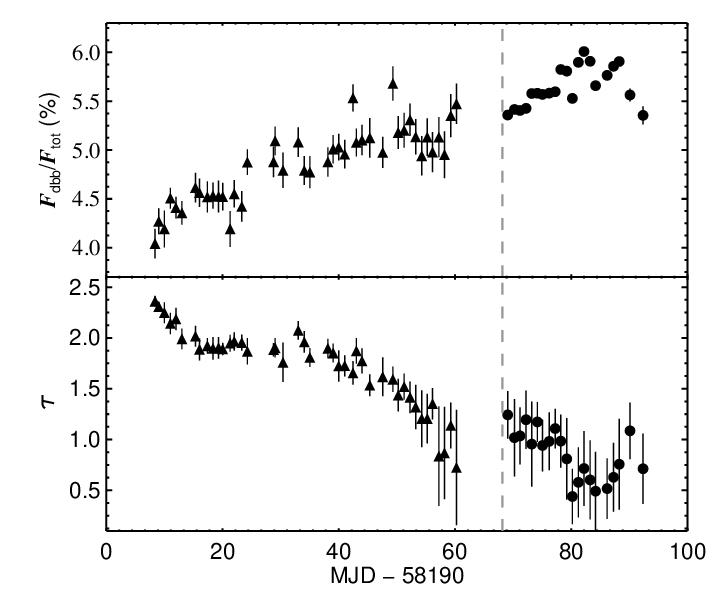}}
\hspace{0.1cm}
\mbox{\includegraphics[width=0.65\linewidth]{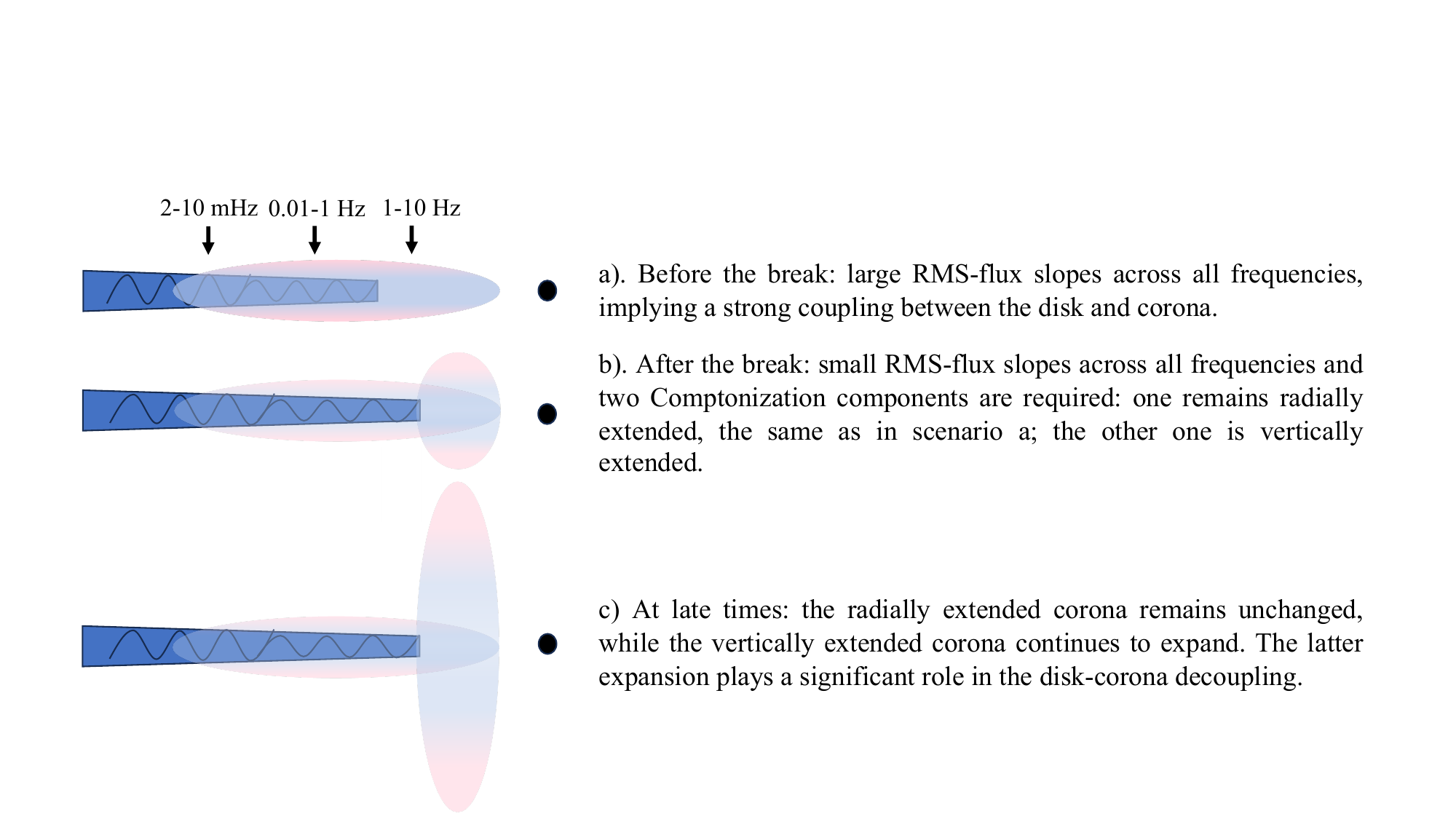}}
\caption{\textbf{Left:} Temporal evolution of the \texttt{blackbody} flux ratio and the optical depth of the corona. \textbf{Right:} Schematics of the disk-corona coupling system. Each spherical region corresponds to a corona, in which the bluish and pinkish region represent the colder and the hotter part, respectively.} 
\label{fig:schematic}
\end{figure*}

To first figure out the origin of the studied emission and their associated evolution of \sub, we adopt the flux of each spectral component from \cite{Wang2020}, and the photon index and the electron temperature of the Comptonization component from \cite{You2021}. 
After calculations, we obtain the ratio of the \texttt{diskbb} flux to the total flux, as well as the optical depth of the Comptonization region. These are presented in the left panel of Fig.~\ref{fig:schematic}, respectively.
It indicates that merely 4--6\% of the emission arises from the accretion disk, implying that the predominant observed emission should come from the Comptonization region, characterized by a declining optical depth. 
In conjunction with the presence of a lognormal distribution of fluxes and the linear RMS-flux relation, these findings provide support for the notion that the observed variability arises from fluctuations generated within the accretion disk and subsequently propagating inwards from the disk to the corona (e.g. \citealt{Lyubarskii1997,Kotov2001,Arevalo2006}).

As further reported by \cite{Wang2020}, they observed a distinct behavior of the time lags at low (2--10\,mHz) and high frequencies (1--10\,Hz). The former presents a random evolution along with the outburst and the latter shows a strong correlation with the spectral index. Consequently, they suggested the former corresponds to viscous propagation of mass accretion fluctuations within the inner regions of the disk and the latter is result of the inverse Comptonization in a jet. 
These results are in line with the fact that the frequency-dependent feature appearing in the power spectrum is connected to the position of the emitting region within accretion systems (e.g. \citealt{Miyamoto1989,Kotov2001,Wilkinson2009,Grinberg2014,Uttley2014,Axelsson2018}). In other words, for features associated with lower/higher frequencies, the emitting region should be more extended/compact and locate further away from/nearer to the central object.Therefore, only the fluctuation with a timescale longer than the diffusion timescale can reach the inner region and the fluctuation produced in the outer disk should be modified by the fluctuation produced in the inner disk.


\subsection{A two-Component corona scenario: radially and vertically extended coronae}

\subsubsection{Scenario assumptions and premises}
Drawing from the prior researches (e.g. \citealt{Lyubarskii1997,Uttley2005}), we present a scenario (see Equations~\ref{equ:flux_rms} to \ref{equ:4}) for the disk-corona coupling to offer a qualitative explanation for the overall evolution of the RMS-flux slope from low to high frequencies. 
Here the low-frequency variability (e.g. 2--10\,mHz), it emerges from a larger radius.
As it propagates inwards, it runs into a different medium at smaller radii in which the fluctuations cannot propagate further, as shown by the much lower RMS-rms slope in the top left panel of Fig.~\ref{fig:slope_ene} (also see the fractional rms at different frequencies in Fig.~3 of \citealt{Wang2020}). Conversely, for the high-frequency variability (e.g. 1--10\,Hz), we propose that it originates from a smaller radius, allowing for its observation with less alteration. 
Given that the majority of the observed emission/variability in \sub originates from the Comptonization region (e.g. a corona) where the hard photons are produced by inverse Compton scattering off the low energy disk photons, a large RMS-flux slope signifies a strong connection or coupling between the disk and the corona and vice verse. 

\subsubsection{Interpretation of the results}
Under this framework, we interpret our results as follows.
Before the line break, $k_{\rm before}$ is relatively high and frequency-dependent while $c_{\rm before}$ is positive and independent of frequency.  
The intercept represents the flux at which the RMS equals zero, indicating the presence of a constant component within the system, devoid of variability (e.g. \citealt{Uttley2001,Gleissner2004}). The similarity observed in the intercept-energy curves at different frequencies suggests that there is an additional constant Comptonization region, which is not strongly connected to the accretion disk and thus has little or non-fluctuations in its flux.

After the line break, there is a notable decrease in both the variability and the RMS-flux slope, spanning from low to high frequencies. 
This implies a weakening of the disk-corona coupling and the domination of the non-fluctuating component in the total flux. 
Since fluctuations within the disk can only propagate radially, a natural outcome of vertical corona expansion is the decoupling of the disk-corona system. This expansion aligns with the decrease in the optical depth of the corona, as shown in the left panel of Fig.~\ref{fig:schematic}.
Hence, we propose that the RMS-flux slope can serve as a metric for quantifying the degree of disk-corona coupling. 
Additionally, the low-frequency slope experiences its initial decline at the highest energies, approximately 36.3\,days before the spectral state transition, as determined by time-averaged broadband spectroscopy. As this decline occurs gradually with decreasing energy, it implies that the hotter portion of the corona initiates its expansion ahead of the colder portion.
Therefore, the deviation in the line break time from the spectral transition time corresponds to the duration of the corona expansion.

On the other hand, an initial negative value for $c_{\rm after}$ is required after the break, indicating the presence of two distinct components, $F_{1}$ and $F_{2}$. We propose that $F_{1}$ is akin to the predominant component before the break, contributing to a linear RMS-flux relation with the same slope, while $F_{2}$ contributes to a constant RMS (as described in Equations~\ref{equ:2}--\ref{equ:4}).
Under this assumption, $F_{1}$ and $F_{2}$ should correspond to a radially and vertically extended corona, respectively, and the decrease in the observed RMS-flux slope is more attributed to $F_{2}$ increase.
To distinguish between these two components, we have introduced a ratio, denoted as $\eta$, which represents the proportion of $F_{1}$ to $F$, and have found that this ratio increases with frequency (as depicted in Fig.~\ref{fig:alpha_f2}). This means that $F_{2}$ exerts less influence at higher frequencies. Therefore, the decoupling between the disk and the vertical corona is less pronounced at lower frequencies/distant regions, but becomes more pronounced at higher frequencies/nearby regions. 
This observation aligns with our corona expansion hypothesis.
As anticipated by our scenario, this expansion of the corona would lead to a diminished reflection component in the spectrum, a phenomenon that has been observed and confirmed by \cite{You2021} (see Fig.~3 in their paper).
Furthermore, the evolution of $\eta$ in relation to energy closely resembles the $k_{\rm after}$-energy curve, as depicted in Figs.~\ref{fig:alpha_f2} and \ref{fig:slope_ene}. As a result, at lower frequencies, where both $\eta$ and $k_{\rm after}$ are low and increase with energy, it implies that the more distant portion of the vertically extended corona should possess a smaller radial extent. Conversely, at higher frequencies, where both $\eta$ and $k_{\rm after}$ are high and exhibit a slight decrease with energy, it suggests that the central part of the vertically extended corona should have a larger radial extent.
Additionally, we observed that the ratio between $F_{1}$ and $F_{2}$ remains constant across different energy bands and frequencies (as shown in Fig.~\ref{fig:alpha_f2}). This observation suggests that the source of seed photons for both the Comptonization components is likely the same, though their dynamical responses to the propagating accretion fluctuations are completely different.

\subsubsection{Caveats}
The assumption that $F_2$ contributes only a constant RMS does not seem completely physical. The motivation of this assumption is to delineate $F_2$ from the observed proportionality between RMS and the total flux $F$. It is possible that $F_2$ does have intrinsic variability independent of the propagating accretion fluctuations, which we simply describe as a ``constant" term on average. The fact that Equation~\ref{equ:4} describes adequately the data after the break justifies our assumption. The ``constant" flux $c_{\rm before}$ obtained from Equation~\ref{equ:flux_rms} also does not seem completely physical. It is reasonable to assume that Equations~\ref{equ:2} and \ref{equ:3} are also true before the break. In this case, the larger slope before the break indicates that $F_1$ dominates before the break and $F_2$ contributes less to the total flux. Therefore we can simply assume that the ``constant" flux term is the average flux of $F_2$ before the break. The fact that $c_{\rm before}$-energy relations exhibit similar profiles at the three different frequency ranges suggests that $F_2$ itself is frequency-independent; this is also consistent with our general picture that the vertically extending corona producing $F_2$ does not respond sensitively to the propagating accretion fluctuations. Since the observed RMS is also dominated by $F_1$ before the break, ignoring the contribution of the ``constant" RMS of $F_2$ before the break is thus a good approximation, though this ``constant" RMS is required after the break due to the dominance of $F_2$ especially at low frequencies.

Moreover, Fig.~\ref{fig:rms_flux_resi} illustrates some excess in the residuals at the high-frequency-high flux end (highlighted by the yellow and blue triangles in the bottom-left panels). This suggests the presence of a third component near the central engine contributing to the additional variability. Given that the studied period is dominated by non-thermal emission, this additional contributor could be the corona and/or jet. Persistent radio emission, considered as evidence of jets, has been detected by the Arcminute Microkelvin Imager Large Array and the Multi-Element Radio Linked Interferometer Network from MJD~58193 to 58300 \citep{Atri2020,Thomas2022}. However, distinguishing the corona and jet emissions from each other is an unresolved matter and is not the primary focus of the current work.

\subsubsection{Scenario Summary}
In summary, our findings suggest the existence of two distinct Comptonization regions within the system: one remains radially extended (i.e. strongly coupled with the disk) and allows for fluctuations to propagate from the disk to the corona; the other region is vertically extended (i.e. weakly coupled with the disk), and continues to expand at late times.
We show three schematics of the overall corona evolution in the right side of Fig.~\ref{fig:schematic}. The further expansion of the vertically extended corona is partially the driving force behind the observed hard-to-hard transition during the bright decaying phase of \sub.

\subsection{Comparison with other studies}
Through the analysis of reverberation lags using \Ni data, authors including \cite{Kara2019}, \cite{Wang2021}, and \cite{DeMarco2021} have examined the disk-corona behavior during the first outburst of \sub, specifically on MJD~58198--58246, MJD~58298--58305, and MJD~58189--58292, respectively.
In these studies, \cite{Kara2019} observed a contracting corona, while \cite{Wang2021} observed an expanding corona, both of which assumes a constant disk inner radius.
Conversely, \cite{DeMarco2021} interpreted the evolution of the frequency related to the reverberation lags as resulting from the reduction in the disk inner radius during MJD~58189--58292, and from the initiation of relativistic jet ejections after MJD~58292.
Our dataset serves to bridge the gap between the data reported by \cite{Kara2019} and \cite{Wang2021}. In contrast to the conclusions drawn by \cite{Kara2019} and \cite{Wang2021}, our findings suggest that corona expansion commenced as early as MJD~58220. This apparent discrepancy is likely attributed to the energy bands used for data analysis. When considering data solely below 10\,keV, the corona expansion appears to initiate around MJD~58240.

The stability of the disk inner radius in \sub has been debated, with some studies suggesting constancy (e.g. \citealt{Kara2019,Buisson2019}) and others indicating a decrease (e.g. \citealt{Wang2020, Zdziarski2021,DeMarco2021}) before the flaring phase of the outburst (around $\sim$MJD~58292). Under either condition, it necessitates the presence of a vertically expanding corona to account for the observed reduction in the RMS-flux slope. In our proposed schematics (referring to Fig.~\ref{fig:schematic}), we opt for the representation that assumes a moving inward accretion disk (adopted from \citealt{Wang2020}). 
The concept of two-component Comptonization being present during the observed period of \sub has previously been suggested by \cite{Zdziarski2021} and \cite{Kawamura2023}. In fact, the post-break coronal geometries we propose bear resemblance to that presented in \cite{Zdziarski2021}. However, in contrast to their work, we offer a dynamic depiction of the evolution of the coronae.
Also, we do not exclude the possibility that the vertically expanding corona may eventually manifest as the radio-detected jet observed at late times.

Additionally, \cite{Ma2023} observed a QPO signal evolving from type C to type B in one \Ni observation of \sub on $\sim$MJD~58305. Combined with the radio flare observed by Arcminute Microkelvin Imager Large Array, they interpreted this transition as indicative of a large-scale corona evolving from a sandwich geometry that covered the inner region of the disk to a vertically extended geometry positioned above the central engine within 1.2\,days. A similar corona configuration has been previously proposed by \cite{Garcia2021, Bellavita2022} to elucidate the properties of the type-B QPOs observed in the BHXRB MAXI~J1348--630.  
Although both \cite{Ma2023} and our work suggested an evolving two-component Componization region for \sub, the studied period and the timescales proposed for the coronae evolution are notably different.



\section{Conclusion}
In order to elucidate the comprehensive evolution of the RMS-flux relation for \sub within the frequency range of 0.002 to 10\,Hz, we propose a scenario featuring two distinct Comptonization components. Our scenario effectively accounts for the dynamic evolution of the RMS-flux slope and intercept across a wide range of frequencies and energy bands. Notably, the concept of a vertically expanding corona is substantiated by the observed decrease in the reflection fraction as determined through the reflection fitting method \citep{You2021}. The pursuit of a more in-depth analysis of the co-evolution of the two-component Comptonization through spectroscopy is hindered by the complexities of isolating and deconstructing the frequency-dependent components present within the spectra.

\section*{Acknowledgements} 
The authors thank the anonymous referee for the constructive comments. YW acknowledges support from the Strategic Priority Research Program of the Chinese Academy of Sciences (Grant No. XDB0550200) and the National Natural Science Foundation of China (Grant No. 12173103). 
SZ acknowledges support from the National Natural Science Foundation of China (Grant No. 12333007 and 12027803) and International Partnership Program of Chinese Academy of Sciences (Grant No. 113111KYSB20190020).
This work made use of the data from the {\hx} mission, a project funded by China National Space Administration (CNSA) and the Chinese Academy of Sciences (CAS). 

\section*{DATA AVAILABILITY}
{\it Insight}-HXMT.

\bibliographystyle{aasjournal}
\bibliography{ref_2019}

\begin{thebibliography}{}
\expandafter\ifx\csname natexlab\endcsname\relax\def\natexlab#1{#1}\fi
\providecommand{\url}[1]{\href{#1}{#1}}

\bibitem[{{Ar{\'e}valo} \& {Uttley}(2006)}]{Arevalo2006}
{Ar{\'e}valo}, P., \& {Uttley}, P. 2006, \mnras, 367, 801

\bibitem[{{Atri} {et~al.}(2020){Atri}, {Miller-Jones}, {Bahramian}, {Plotkin},
  {Deller}, {Jonker}, {Maccarone}, {Sivakoff}, {Soria}, {Altamirano},
  {Belloni}, {Fender}, {Koerding}, {Maitra}, {Markoff}, {Migliari}, {Russell},
  {Russell}, {Sarazin}, {Tetarenko}, \& {Tudose}}]{Atri2020}
{Atri}, P., {Miller-Jones}, J.~C.~A., {Bahramian}, A., {et~al.} 2020, \mnras,
  493, L81

\bibitem[{{Axelsson} \& {Done}(2018)}]{Axelsson2018}
{Axelsson}, M., \& {Done}, C. 2018, \mnras, 480, 751

\bibitem[{{Bassi} {et~al.}(2019){Bassi}, {Del Santo}, {D'A{\i}}, {Motta},
  {Malzac}, {Segreto}, {Miller-Jones}, {Atri}, {Plotkin}, {Belloni}, {Mineo},
  \& {Tzioumis}}]{Bassi2019}
{Bassi}, T., {Del Santo}, M., {D'A{\i}}, A., {et~al.} 2019, \mnras, 482, 1587

\bibitem[{{Bellavita} {et~al.}(2022){Bellavita}, {Garc{\'\i}a}, {M{\'e}ndez},
  \& {Karpouzas}}]{Bellavita2022}
{Bellavita}, C., {Garc{\'\i}a}, F., {M{\'e}ndez}, M., \& {Karpouzas}, K. 2022,
  \mnras, 515, 2099

\bibitem[{{Buisson} {et~al.}(2019){Buisson}, {Fabian}, {Barret}, {F{\"u}rst},
  {Gandhi}, {Garc{\'\i}a}, {Kara}, {Madsen}, {Miller}, {Parker}, {Shaw},
  {Tomsick}, \& {Walton}}]{Buisson2019}
{Buisson}, D.~J.~K., {Fabian}, A.~C., {Barret}, D., {et~al.} 2019, \mnras, 490,
  1350

\bibitem[{{Capitanio} {et~al.}(2009){Capitanio}, {Belloni}, {Del Santo}, \&
  {Ubertini}}]{Capitanio2009}
{Capitanio}, F., {Belloni}, T., {Del Santo}, M., \& {Ubertini}, P. 2009,
  \mnras, 398, 1194

\bibitem[{{Churazov} {et~al.}(2001){Churazov}, {Gilfanov}, \&
  {Revnivtsev}}]{Churazov2001}
{Churazov}, E., {Gilfanov}, M., \& {Revnivtsev}, M. 2001, \mnras, 321, 759

\bibitem[{{De Marco} {et~al.}(2021){De Marco}, {Zdziarski}, {Ponti},
  {Migliori}, {Belloni}, {Segovia Otero}, {Dzie{\l}ak}, \& {Lai}}]{DeMarco2021}
{De Marco}, B., {Zdziarski}, A.~A., {Ponti}, G., {et~al.} 2021, \aap, 654, A14

\bibitem[{{Garc{\'\i}a} {et~al.}(2021){Garc{\'\i}a}, {M{\'e}ndez}, {Karpouzas},
  {Belloni}, {Zhang}, \& {Altamirano}}]{Garcia2021}
{Garc{\'\i}a}, F., {M{\'e}ndez}, M., {Karpouzas}, K., {et~al.} 2021, \mnras,
  501, 3173

\bibitem[{{Gleissner} {et~al.}(2004){Gleissner}, {Wilms}, {Pottschmidt},
  {Uttley}, {Nowak}, \& {Staubert}}]{Gleissner2004}
{Gleissner}, T., {Wilms}, J., {Pottschmidt}, K., {et~al.} 2004, \aap, 414, 1091

\bibitem[{{Grinberg} {et~al.}(2014){Grinberg}, {Pottschmidt}, {B{\"o}ck},
  {Schmid}, {Nowak}, {Uttley}, {Tomsick}, {Rodriguez}, {Hell}, {Markowitz},
  {Bodaghee}, {Cadolle Bel}, {Rothschild}, \& {Wilms}}]{Grinberg2014}
{Grinberg}, V., {Pottschmidt}, K., {B{\"o}ck}, M., {et~al.} 2014, \aap, 565, A1

\bibitem[{{Heil} {et~al.}(2012){Heil}, {Vaughan}, \& {Uttley}}]{Heil2012}
{Heil}, L.~M., {Vaughan}, S., \& {Uttley}, P. 2012, \mnras, 422, 2620

\bibitem[{{Kara} {et~al.}(2019){Kara}, {Steiner}, {Fabian}, {Cackett},
  {Uttley}, {Remillard}, {Gendreau}, {Arzoumanian}, {Altamirano}, {Eikenberry},
  {Enoto}, {Homan}, {Neilsen}, \& {Stevens}}]{Kara2019}
{Kara}, E., {Steiner}, J.~F., {Fabian}, A.~C., {et~al.} 2019, \nat, 565, 198

\bibitem[{{Kawamura} {et~al.}(2023){Kawamura}, {Done}, {Axelsson}, \&
  {Takahashi}}]{Kawamura2023}
{Kawamura}, T., {Done}, C., {Axelsson}, M., \& {Takahashi}, T. 2023, \mnras,
  519, 4434

\bibitem[{{Kotov} {et~al.}(2001){Kotov}, {Churazov}, \& {Gilfanov}}]{Kotov2001}
{Kotov}, O., {Churazov}, E., \& {Gilfanov}, M. 2001, \mnras, 327, 799

\bibitem[{{Lyubarskii}(1997)}]{Lyubarskii1997}
{Lyubarskii}, Y.~E. 1997, \mnras, 292, 679

\bibitem[{{Ma} {et~al.}(2023){Ma}, {M{\'e}ndez}, {Garc{\'\i}a}, {Sai}, {Zhang},
  \& {Zhang}}]{Ma2023}
{Ma}, R., {M{\'e}ndez}, M., {Garc{\'\i}a}, F., {et~al.} 2023, \mnras, 525, 854

\bibitem[{{Ma} {et~al.}(2021){Ma}, {Tao}, {Zhang}, {Zhang}, {Bu}, {Ge}, {Chen},
  {Qu}, {Zhang}, {Lu}, {Song}, {Yang}, {Yuan}, {Cai}, {Cao}, {Chang}, {Chen},
  {Chen}, {Chen}, {Chen}, {Chen}, {Cui}, {Cui}, {Deng}, {Dong}, {Du}, {Fu},
  {Gao}, {Gao}, {Gao}, {Gu}, {Guan}, {Guo}, {Han}, {Huang}, {Huo}, {Ji}, {Jia},
  {Jiang}, {Jiang}, {Jin}, {Jin}, {Kong}, {Li}, {Li}, {Li}, {Li}, {Li}, {Li},
  {Li}, {Li}, {Li}, {Li}, {Li}, {Liang}, {Liao}, {Liu}, {Liu}, {Liu}, {Liu},
  {Liu}, {Liu}, {Lu}, {Lu}, {Luo}, {Luo}, {Meng}, {Nang}, {Nie}, {Ou}, {Sai},
  {Shang}, {Song}, {Sun}, {Tan}, {Tuo}, {Wang}, {Wang}, {Wang}, {Wang}, {Wang},
  {Wang}, {Wen}, {Wu}, {Wu}, {Wu}, {Xiao}, {Xiao}, {Xie}, {Xiong}, {Xu}, {Xu},
  {Yang}, {Yang}, {Yang}, {Yi}, {Yin}, {You}, {Zhang}, {Zhang}, {Zhang},
  {Zhang}, {Zhang}, {Zhang}, {Zhang}, {Zhang}, {Zhang}, {Zhang}, {Zhang},
  {Zhang}, {Zhang}, {Zhang}, {Zhang}, {Zhang}, {Zhao}, {Zhao}, {Zheng}, {Zhou},
  {Zhou}, {Zhu}, {Zhu}, \& {Zhuang}}]{Ma2021}
{Ma}, X., {Tao}, L., {Zhang}, S.-N., {et~al.} 2021, Nature Astronomy, 5, 94

\bibitem[{{Matsuoka} {et~al.}(2009){Matsuoka}, {Kawasaki}, {Ueno}, {Tomida},
  {Kohama}, {Suzuki}, {Adachi}, {Ishikawa}, {Mihara}, {Sugizaki}, {Isobe},
  {Nakagawa}, {Tsunemi}, {Miyata}, {Kawai}, {Kataoka}, {Morii}, {Yoshida},
  {Negoro}, {Nakajima}, {Ueda}, {Chujo}, {Yamaoka}, {Yamazaki}, {Nakahira},
  {You}, {Ishiwata}, {Miyoshi}, {Eguchi}, {Hiroi}, {Katayama}, \&
  {Ebisawa}}]{Matsuoka2009}
{Matsuoka}, M., {Kawasaki}, K., {Ueno}, S., {et~al.} 2009, \pasj, 61, 999

\bibitem[{{M{\'e}ndez} {et~al.}(1997){M{\'e}ndez}, {van der Klis}, {van
  Paradijs}, {Lewin}, {Lamb}, {Vaughan}, {Kuulkers}, \&
  {Psaltis}}]{Mendez1997a}
{M{\'e}ndez}, M., {van der Klis}, M., {van Paradijs}, J., {et~al.} 1997, \apjl,
  485, L37

\bibitem[{{Miyamoto} \& {Kitamoto}(1989)}]{Miyamoto1989}
{Miyamoto}, S., \& {Kitamoto}, S. 1989, \nat, 342, 773

\bibitem[{{Mu{\~n}oz-Darias} {et~al.}(2011){Mu{\~n}oz-Darias}, {Motta}, \&
  {Belloni}}]{Munoz2011}
{Mu{\~n}oz-Darias}, T., {Motta}, S., \& {Belloni}, T.~M. 2011, \mnras, 410, 679

\bibitem[{{Mushtukov} {et~al.}(2018){Mushtukov}, {Ingram}, \& {van der
  Klis}}]{Mushtukov2018}
{Mushtukov}, A.~A., {Ingram}, A., \& {van der Klis}, M. 2018, \mnras, 474, 2259

\bibitem[{{Remillard} \& {McClintock}(2006)}]{Remillard2006}
{Remillard}, R.~A., \& {McClintock}, J.~E. 2006, \araa, 44, 49

\bibitem[{{Stiele} \& {Kong}(2020)}]{Stiele2020}
{Stiele}, H., \& {Kong}, A.~K.~H. 2020, \apj, 889, 142

\bibitem[{{Tanaka} \& {Lewin}(1995)}]{Tanaka1995}
{Tanaka}, Y., \& {Lewin}, W.~H.~G. 1995, in X-ray Binaries, 126--174

\bibitem[{{Thomas} {et~al.}(2022){Thomas}, {Charles}, {Buckley}, {Kotze},
  {Lasota}, {Potter}, {Steiner}, \& {Paice}}]{Thomas2022}
{Thomas}, J.~K., {Charles}, P.~A., {Buckley}, D. A.~H., {et~al.} 2022, \mnras,
  509, 1062

\bibitem[{{Uttley} {et~al.}(2014){Uttley}, {Cackett}, {Fabian}, {Kara}, \&
  {Wilkins}}]{Uttley2014}
{Uttley}, P., {Cackett}, E.~M., {Fabian}, A.~C., {Kara}, E., \& {Wilkins},
  D.~R. 2014, \aapr, 22, 72

\bibitem[{{Uttley} \& {McHardy}(2001)}]{Uttley2001}
{Uttley}, P., \& {McHardy}, I.~M. 2001, \mnras, 323, L26

\bibitem[{{Uttley} {et~al.}(2005){Uttley}, {McHardy}, \&
  {Vaughan}}]{Uttley2005}
{Uttley}, P., {McHardy}, I.~M., \& {Vaughan}, S. 2005, \mnras, 359, 345

\bibitem[{{van der Klis}(1995)}]{Klis1995}
{van der Klis}, M. 1995, in X-ray Binaries, 252--307

\bibitem[{{Wang} {et~al.}(2021){Wang}, {Mastroserio}, {Kara}, {Garc{\'\i}a},
  {Ingram}, {Connors}, {van der Klis}, {Dauser}, {Steiner}, {Buisson}, {Homan},
  {Lucchini}, {Fabian}, {Bright}, {Fender}, {Cackett}, \&
  {Remillard}}]{Wang2021}
{Wang}, J., {Mastroserio}, G., {Kara}, E., {et~al.} 2021, \apjl, 910, L3

\bibitem[{{Wang} {et~al.}(2020){Wang}, {Ji}, {Zhang}, {M{\'e}ndez}, {Qu},
  {Maggi}, {Ge}, {Qiao}, {Tao}, {Zhang}, {Altamirano}, {Zhang}, {Ma}, {Lu},
  {Li}, {Huang}, {Zheng}, {Chen}, {Chang}, {Tuo}, {G{\"u}ng{\"o}r}, {Song},
  {Xu}, {Cao}, {Chen}, {Liu}, {Bu}, {Cai}, {Chen}, {Chen}, {Chen}, {Chen},
  {Cui}, {Cui}, {Deng}, {Dong}, {Du}, {Fu}, {Gao}, {Gao}, {Gao}, {Gu}, {Guan},
  {Guo}, {Han}, {Huo}, {Jia}, {Jiang}, {Jiang}, {Jin}, {Jin}, {Kong}, {Li},
  {Li}, {Li}, {Li}, {Li}, {Li}, {Li}, {Li}, {Li}, {Li}, {Liang}, {Liao}, {Liu},
  {Liu}, {Liu}, {Liu}, {Lu}, {Lu}, {Luo}, {Luo}, {Meng}, {Nang}, {Nie}, {Ou},
  {Sai}, {Shang}, {Song}, {Sun}, {Tan}, {Wang}, {Wang}, {Wang}, {Wang}, {Wang},
  {Wen}, {Wu}, {Wu}, {Wu}, {Xiao}, {Xiao}, {Xiong}, {Yang}, {Yang}, {Yang},
  {Yang}, {Yi}, {Yin}, {You}, {Zhang}, {Zhang}, {Zhang}, {Zhang}, {Zhang},
  {Zhang}, {Zhang}, {Zhang}, {Zhang}, {Zhang}, {Zhang}, {Zhang}, {Zhang},
  {Zhang}, {Zhang}, {Zhang}, {Zhao}, {Zhao}, {Zhou}, {Zhou}, {Zhuang}, {Zhu},
  {Zhu}, \& {Wang}}]{Wang2020}
{Wang}, Y., {Ji}, L., {Zhang}, S.~N., {et~al.} 2020, \apj, 896, 33

\bibitem[{{Wilkinson} \& {Uttley}(2009)}]{Wilkinson2009}
{Wilkinson}, T., \& {Uttley}, P. 2009, \mnras, 397, 666

\bibitem[{{You} {et~al.}(2021){You}, {Tuo}, {Li}, {Wang}, {Zhang}, {Zhang},
  {Ge}, {Luo}, {Liu}, {Yuan}, {Dai}, {Liu}, {Qiao}, {Jin}, {Liu}, {Czerny},
  {Wu}, {Bu}, {Cai}, {Cao}, {Chang}, {Chen}, {Chen}, {Chen}, {Chen}, {Chen},
  {Chen}, {Cui}, {Cui}, {Deng}, {Dong}, {Du}, {Fu}, {Gao}, {Gao}, {Gao}, {Gu},
  {Guan}, {Guo}, {Han}, {Huang}, {Huo}, {Jia}, {Jiang}, {Jiang}, {Jin}, {Jin},
  {Kong}, {Li}, {Li}, {Li}, {Li}, {Li}, {Li}, {Li}, {Li}, {Li}, {Li}, {Li},
  {Liang}, {Liao}, {Liu}, {Liu}, {Liu}, {Liu}, {Liu}, {Lu}, {Lu}, {Lu}, {Luo},
  {Luo}, {Ma}, {Meng}, {Nang}, {Nie}, {Ou}, {Qu}, {Sai}, {Shang}, {Song},
  {Song}, {Sun}, {Tan}, {Tao}, {Wang}, {Wang}, {Wang}, {Wang}, {Wang}, {Wang},
  {Wen}, {Wu}, {Wu}, {Wu}, {Xiao}, {Xiao}, {Xiong}, {Xu}, {Yang}, {Yang},
  {Yang}, {Yi}, {Yin}, {You}, {Zhang}, {Zhang}, {Zhang}, {Zhang}, {Zhang},
  {Zhang}, {Zhang}, {Zhang}, {Zhang}, {Zhang}, {Zhang}, {Zhang}, {Zhang},
  {Zhang}, {Zhang}, {Zhao}, {Zhao}, {Zheng}, {Zhou}, {Zhou}, {Zhu}, \&
  {Zhu}}]{You2021}
{You}, B., {Tuo}, Y., {Li}, C., {et~al.} 2021, Nature Communications, 12, 1025

\bibitem[{{Zdziarski} {et~al.}(2021){Zdziarski}, {Dzie{\l}ak}, {De Marco},
  {Szanecki}, \& {Nied{\'z}wiecki}}]{Zdziarski2021}
{Zdziarski}, A.~A., {Dzie{\l}ak}, M.~A., {De Marco}, B., {Szanecki}, M., \&
  {Nied{\'z}wiecki}, A. 2021, \apjl, 909, L9

\end{thebibliography}

\appendix
\renewcommand{\thefigure}{\hbAppendixPrefix\arabic{figure}}
\numberwithin{figure}{section}
\section{Rms-$\eta$ relation under different assumptions} 
\begin{figure*}  
\centering
\mbox{\includegraphics[width=3.3cm]{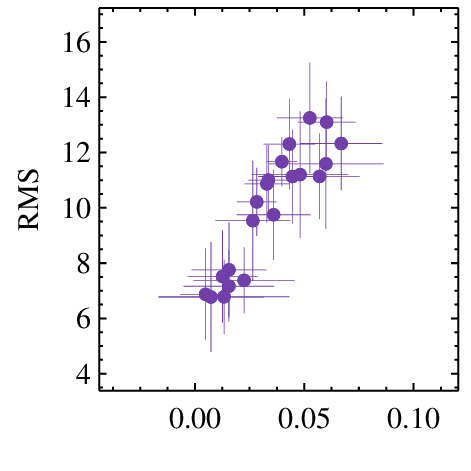}}
\hspace{-0.6cm}
\vspace{-0.2cm}
\mbox{\includegraphics[width=3.3cm]{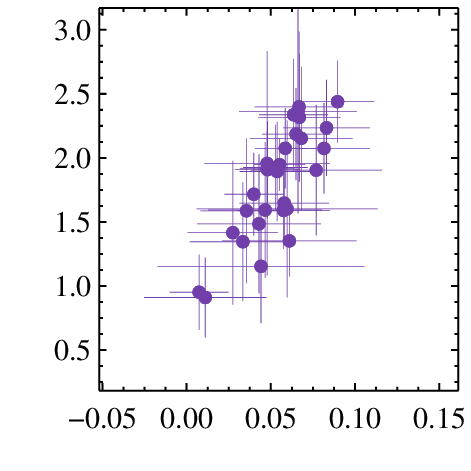}}
\hspace{-0.6cm}
\mbox{\includegraphics[width=3.3cm]{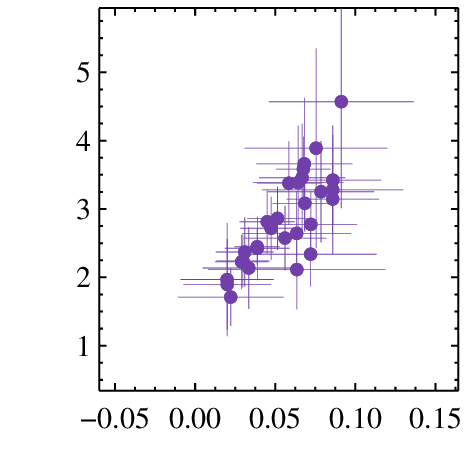}}
\hspace{-0.6cm}
\mbox{\includegraphics[width=3.3cm]{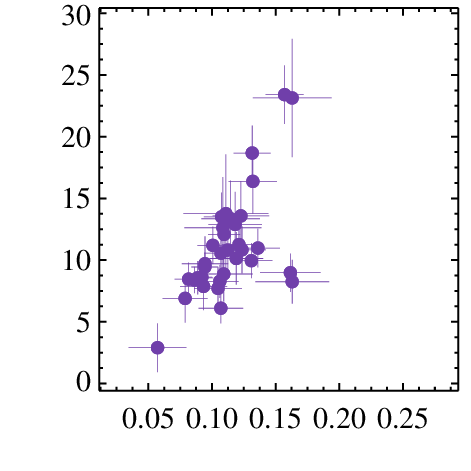}}
\hspace{-0.6cm}
\mbox{\includegraphics[width=3.3cm]{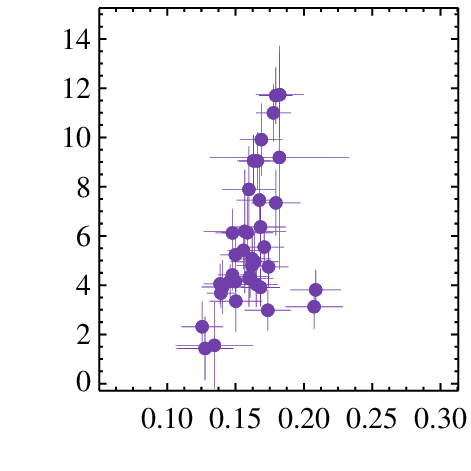}}
\hspace{-0.6cm}
\mbox{\includegraphics[width=3.3cm]{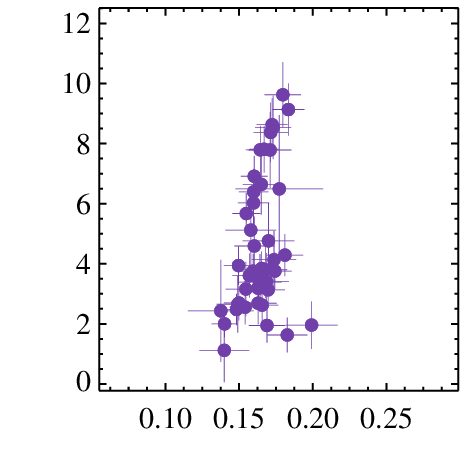}}
\vspace{-0.2cm}
\mbox{\includegraphics[width=3.3cm]{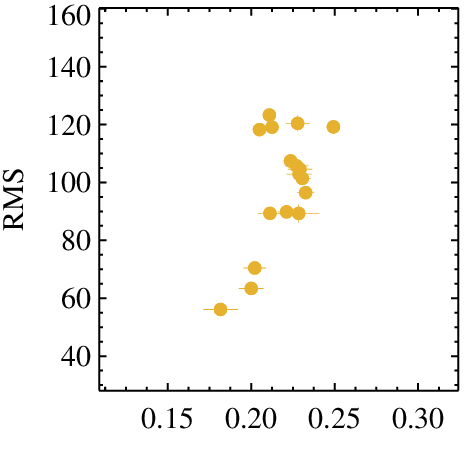}}
\hspace{-0.6cm}
\mbox{\includegraphics[width=3.3cm]{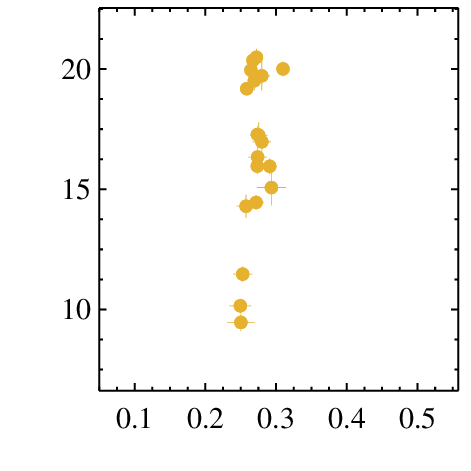}}
\hspace{-0.6cm}
\mbox{\includegraphics[width=3.3cm]{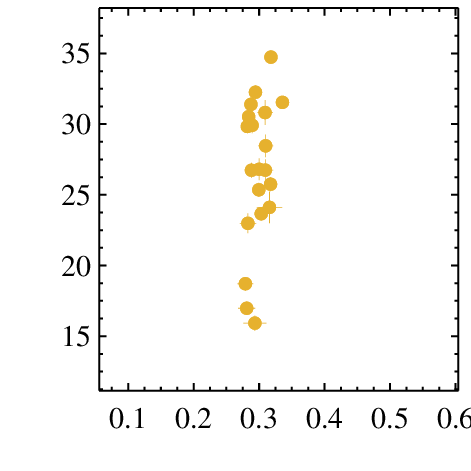}}
\hspace{-0.6cm}
\mbox{\includegraphics[width=3.3cm]{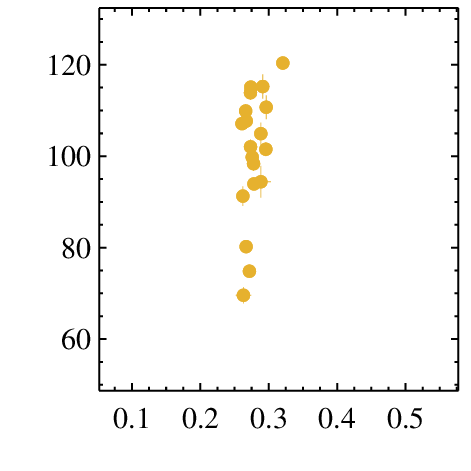}}
\hspace{-0.6cm}
\mbox{\includegraphics[width=3.3cm]{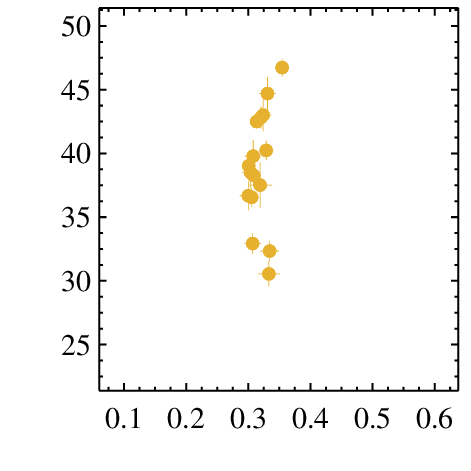}}
\hspace{-0.6cm}
\mbox{\includegraphics[width=3.3cm]{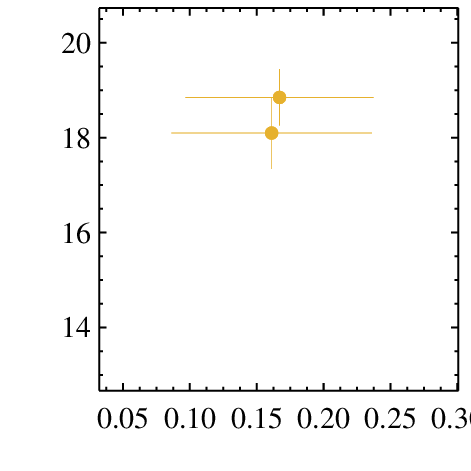}}
\mbox{\includegraphics[width=3.3cm]{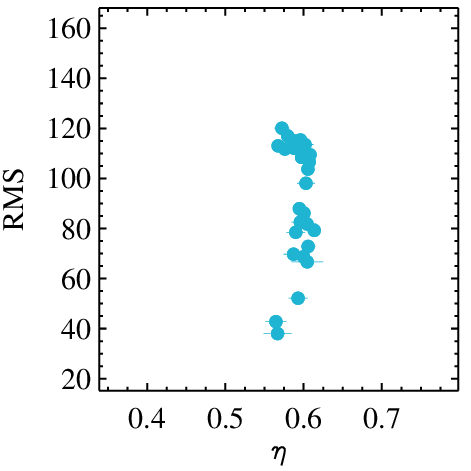}}
\hspace{-0.6cm}
\mbox{\includegraphics[width=3.3cm]{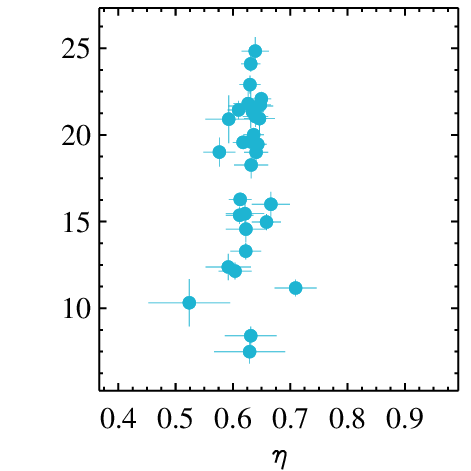}}
\hspace{-0.6cm}
\mbox{\includegraphics[width=3.3cm]{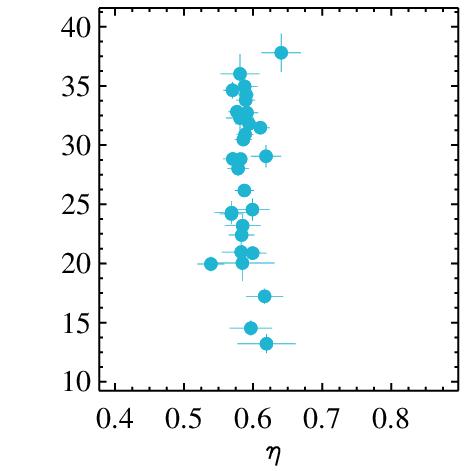}}
\hspace{-0.6cm}
\mbox{\includegraphics[width=3.3cm]{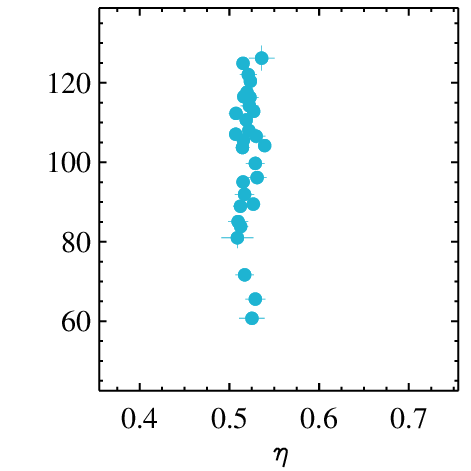}}
\hspace{-0.6cm}
\mbox{\includegraphics[width=3.3cm]{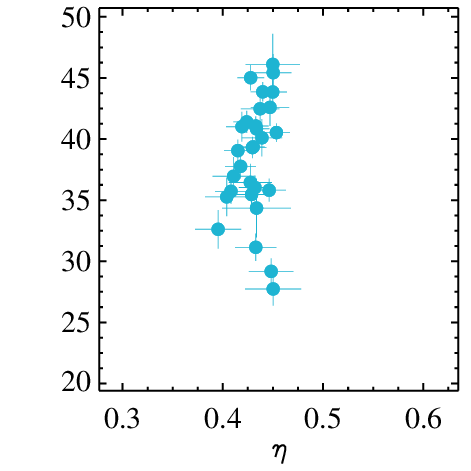}}
\hspace{-0.6cm}
\mbox{\includegraphics[width=3.3cm]{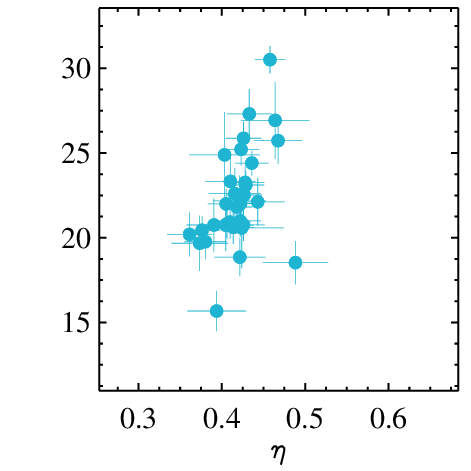}}
\vspace{-0.1cm}
\caption{Energy-dependent RMS-$\eta$ relation in {\sub} in three frequency ranges when using Equation~\ref{equ:4}. From top to bottom, frequency increases while from left to right, energy increases.
}
\label{fig:rms_eta}
\end{figure*}

As presented in Section~\ref{sec:result_rms_flux}, we introduced a second component to address the negative intercept issue and employed a ratio, $\eta=F_{1}/F$, to characterize these two components. The relationship between RMS and $\eta$ at various frequencies and energies is depicted in Fig.~\ref{fig:rms_eta}. Notably, with the exception of cases involving low frequencies and low energies, $\eta$ remains largely independent of RMS (as well as flux).

\begin{figure*}  
\centering
\mbox{\includegraphics[width=3.3cm]{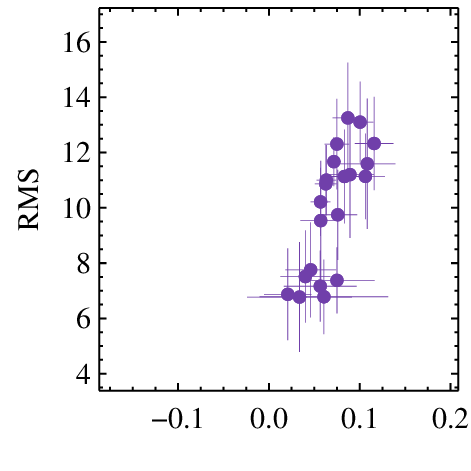}}
\hspace{-0.6cm}
\vspace{-0.2cm}
\mbox{\includegraphics[width=3.3cm]{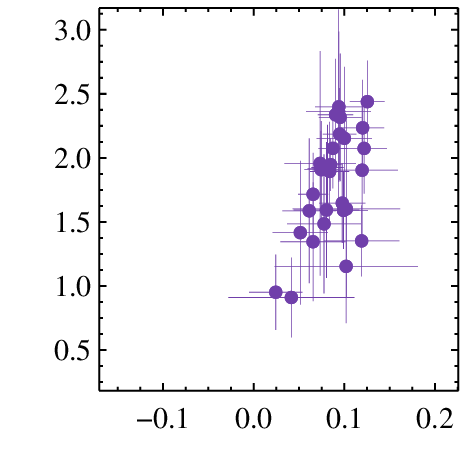}}
\hspace{-0.6cm}
\mbox{\includegraphics[width=3.3cm]{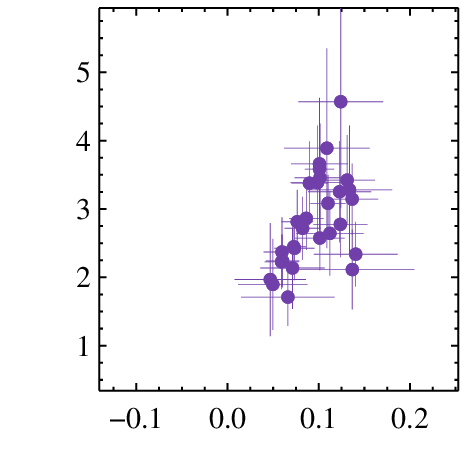}}
\hspace{-0.6cm}
\mbox{\includegraphics[width=3.3cm]{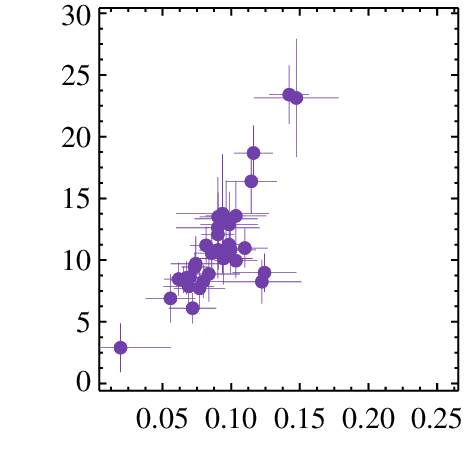}}
\hspace{-0.6cm}
\mbox{\includegraphics[width=3.3cm]{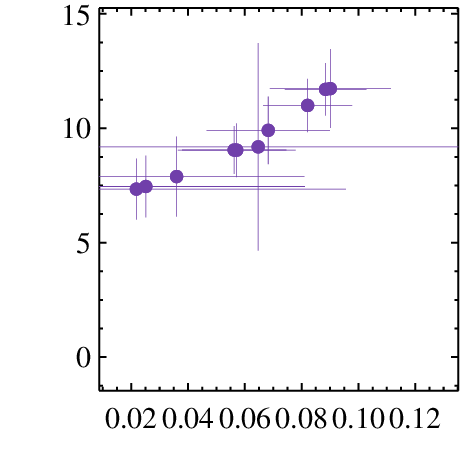}}
\hspace{-0.6cm}
\mbox{\includegraphics[width=3.3cm]{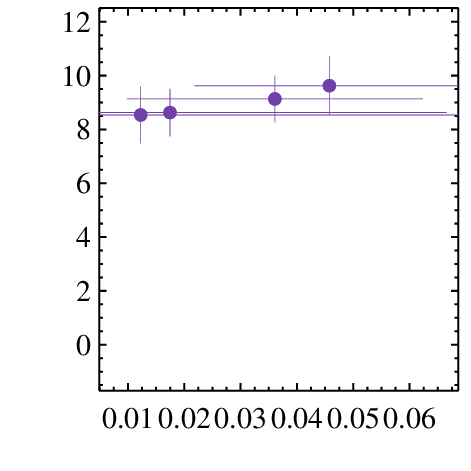}}
\vspace{-0.2cm}
\mbox{\includegraphics[width=3.3cm]{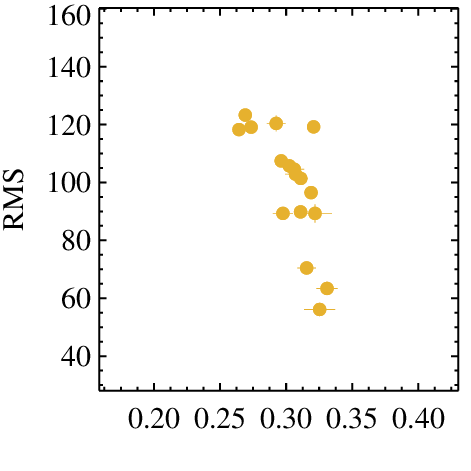}}
\hspace{-0.6cm}
\mbox{\includegraphics[width=3.3cm]{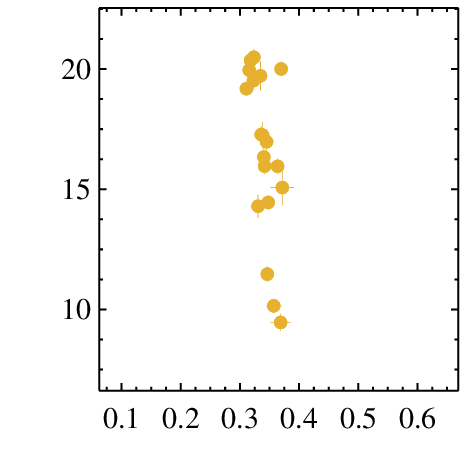}}
\hspace{-0.6cm}
\mbox{\includegraphics[width=3.3cm]{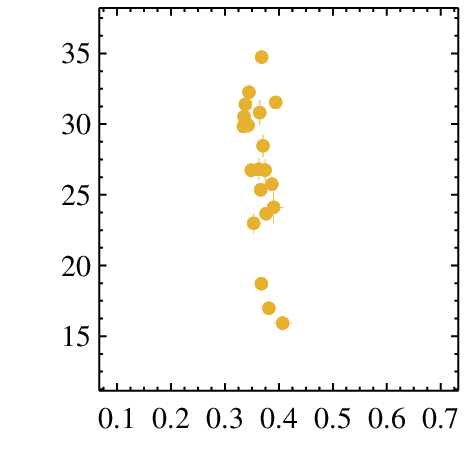}}
\hspace{-0.6cm}
\mbox{\includegraphics[width=3.3cm]{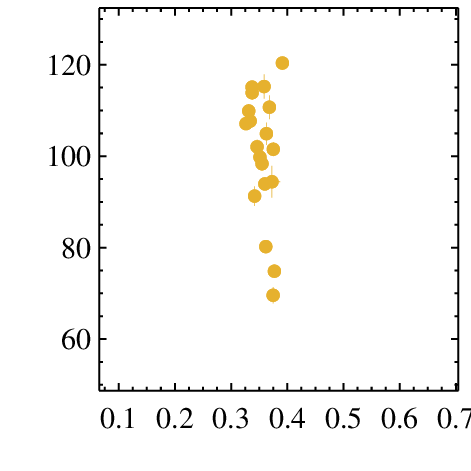}}
\hspace{-0.6cm}
\mbox{\includegraphics[width=3.3cm]{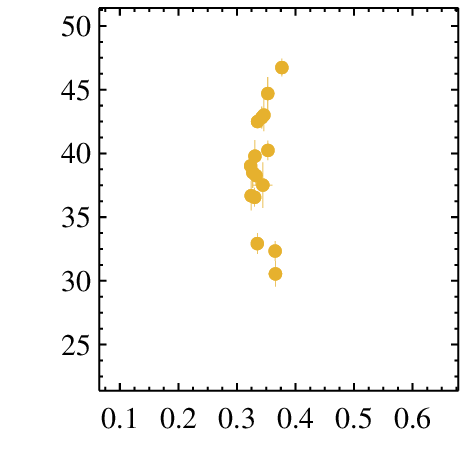}}
\hspace{-0.6cm}
\mbox{\includegraphics[width=3.3cm]{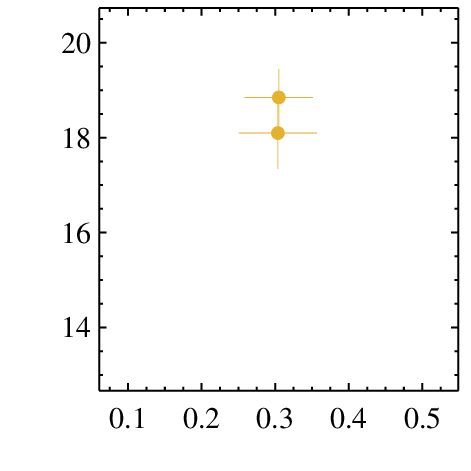}}
\vspace{-0.2cm}
\mbox{\includegraphics[width=3.3cm]{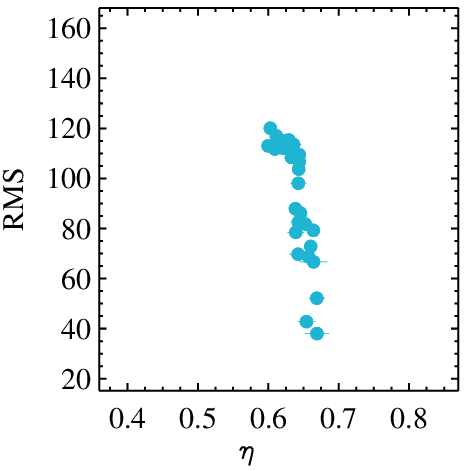}}
\hspace{-0.6cm}
\mbox{\includegraphics[width=3.3cm]{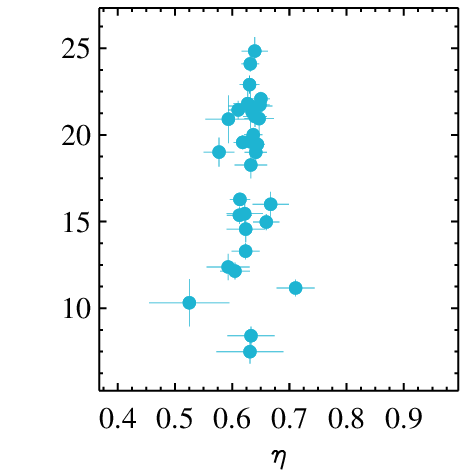}}
\hspace{-0.6cm}
\mbox{\includegraphics[width=3.3cm]{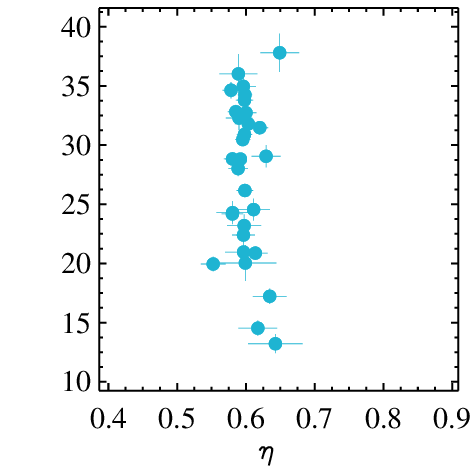}}
\hspace{-0.6cm}
\mbox{\includegraphics[width=3.3cm]{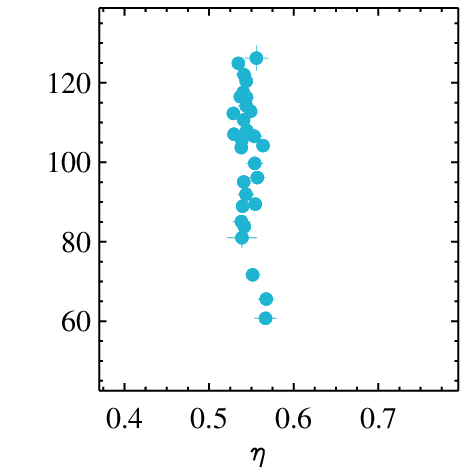}}
\hspace{-0.6cm}
\mbox{\includegraphics[width=3.3cm]{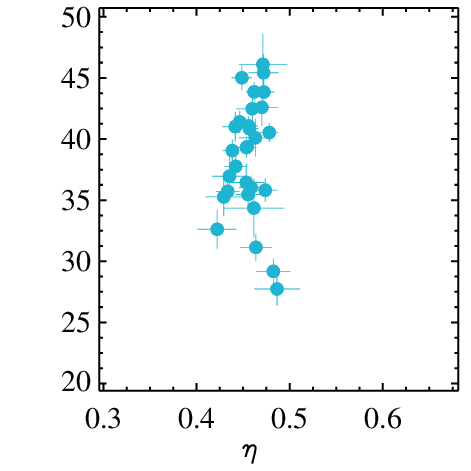}}
\hspace{-0.6cm}
\mbox{\includegraphics[width=3.3cm]{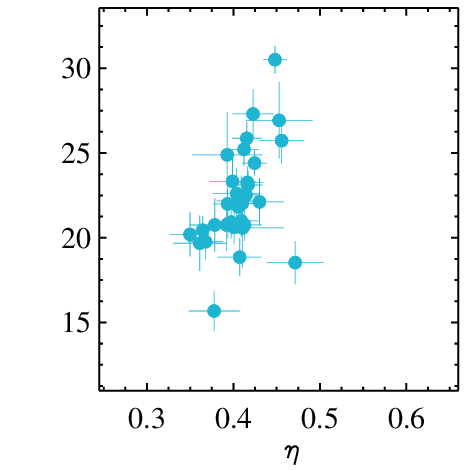}}
\vspace{-0.1cm}
\caption{Energy-dependent RMS-$\eta$ relation in {\sub} in three frequency ranges when using Equation~\ref{equ:5}. From top to bottom, frequency increases while from left to right, energy increases.
}
\label{fig:rms_eta2}
\end{figure*}

Furthermore, in an attempt to account for the influence of independent stochastic fluctuations, we explored an alternative formulation of Equation~\ref{equ:4}, namely, 
\begin{equation}
 \sigma(t)^2=[k_{\rm before}*F_{1}(t)]^2+a^2. \label{equ:5}
\end{equation}
It is important to note that the new equation is only applicable when $\sigma(t)^2 \geq k_{\rm before}$, and accordingly, we present the data that meets this criterion in Fig.~\ref{fig:rms_eta2}. 
Eventually, we found that this modified equation yields comparable results for $\eta$ at low and high frequencies from both equations. While in the frequency range of 0.01--1\,Hz, the newly obtained $\eta$ tends to be relatively higher compared to the original value. Nevertheless, its association with RMS or flux remains consistent.

\end{document}